\documentclass[lettersize,journal]{IEEEtran}
\usepackage{amsmath,amsfonts}
\usepackage{algorithmic}
\usepackage{algorithm}
\usepackage{array}
\usepackage[caption=false,font=normalsize,labelfont=sf,textfont=sf]{subfig}
\usepackage{textcomp}
\usepackage{stfloats}
\usepackage{url}
\usepackage{verbatim}
\usepackage{graphicx}
\usepackage{cite}
\usepackage{xcolor}
\usepackage{mathtools}
\usepackage{amssymb}
\usepackage{balance}
\usepackage{multicol}
\usepackage{multirow}

\begin{document}

\title{A Survey of Attacks on Large Language Models}

\author{Wenrui Xu,~\IEEEmembership{Graduate Student Member,~IEEE} and Keshab K. Parhi,~\IEEEmembership{Life Fellow,~IEEE}
\thanks{The authors are with the Department of Electrical and Computer Engineering, University of Minnesota, Minneapolis, MN 55455 USA (e-mail:
xu000424@umn.edu; parhi@umn.edu).}
}



\maketitle

\begin{abstract}
Large language models (LLMs) and LLM-based agents have been widely deployed in a wide range of applications in the real world, including healthcare diagnostics, financial analysis, customer support, robotics, and autonomous driving, expanding their powerful capability of understanding, reasoning, and generating natural languages. However, the wide deployment of LLM-based applications exposes critical security and reliability risks, such as the potential for malicious misuse, privacy leakage, and service disruption that weaken user trust and undermine societal safety. This paper provides a systematic overview of the details of adversarial attacks targeting both LLMs and LLM-based agents. These attacks are organized into three phases in LLMs: Training-Phase Attacks, Inference-Phase Attacks, and Availability \& Integrity Attacks. For each phase, we analyze the details of representative and recently introduced attack methods along with their corresponding defenses. We hope our survey will provide a good tutorial and a comprehensive understanding of LLM security, especially for attacks on LLMs. We desire to raise attention to the risks inherent in widely deployed LLM-based applications and highlight the urgent need for robust mitigation strategies for evolving threats.
\end{abstract}


\section{Introduction}
The large language model (LLM) has shown great advancements in recent years; it has become a popular topic of discussion and application in both academic and industrial fields. LLMs, characterized by the massive parameter size, are designed for handling a wide range of natural language processing (NLP) tasks, including text generation~\cite{li2024pre}, question reasoning~\cite{plaat2024reasoning}, and sentiment analysis~\cite{zhang2023sentiment}. Benefiting from the training on the vast amount of text data, the LLMs are capable of understanding and processing human language effectively, enabling them to perform complex language-related tasks accurately. Numerous LLMs such as ChatGPT~\cite{openai2023chatgpt} from OpenAI, LLaMA~\cite{dubey2024llama} from Meta, DeepSeek-R1~\cite{guo2025deepseek} from Deepseek, Grok 3~\cite{xai2025grok3} from xAI were developed and released since 2025; These models are significant milestones in the field of Artificial Intelligence and have gained widespread public attention due to their advanced capability and application in various domains. 

The main features of LLMs~\cite{10.1145/3649506} can be summarized as follows: 1) generalization ability for deep understanding of natural language context; 2) capability of high-quality text generation in a human manner; 3) ability to handle knowledge-intensive tasks; 4) reasoning capability to enhance the process of decision-making and problem-solving. Training strong performance LLMs with achieving these features, normally the vast amount of high-quality training data and large parameter size of LLMs are required by following the scaling laws. 
LLM-based agents, LLM-based autonomous agents~\cite{wang2024survey}, are autonomous systems that utilize the human-like capability of LLMs to execute diverse tasks effectively. It can take well-informed actions without domain-specific training compared to reinforcement learning (RL). The human interaction based on natural language interfaces provided by LLM-based agents might be more flexible and explainable.
However, with the strong capability of LLMs and LLM-based agents in understanding and processing natural languages, the risk associated with various security threats, such as jailbreaking, backdoor attacks, prompt injection, and Denial of Service (DoS) becomes critical and demands more attention. 

With the rising concerns over the security of LLMs, researchers are focused on identifying potential threat models and developing defense strategies related to them. The battle between threat models and defense strategies can be viewed as an arms race between the arrow and the shield. In this paper, we primarily focus on the development and recent advancements in various attack strategies targeting LLMs and LLM-based agents, including the methodology, implications, and challenges posed to LLM security.

This paper provides a comprehensive summary of the development and recent advancements in adversarial attacks on LLMs, including threat strategies such as jailbreaking, backdoor and data poisoning, prompt injection, DoS, and watermarking attacks. In this paper, these attacks are systematically summarized into three different categories: Training-Phase attacks, Inference-Phase attacks, and Availability \& Integrity attacks. Additionally, this paper extends the discussion to the attacks specific to LLM-based agents and highlights the vulnerabilities introduced by the architecture of the agent systems and their interactions with external tools and environments.

The paper is organized as follows: Section \ref{sec:Background} introduces the background of LLMs and LLM-based agents. Section \ref{sec:Overview} presents an overview of attacks discussed in this paper. Section \ref{sec:Train-Phase} provides a summary of Training-Phase attacks, specifically backdoor attacks, on LLM and LLM-based agents. Section \ref{sec:Inference-Phase} illustrates the development of Inference-Phase attacks on LLMs including jailbreaking and prompt Injection. Section \ref{sec: Other} reviews the Availability \& Integrity attacks on the LLMs such as DoS and watermarking attacks.

\section{Background}\label{sec:Background}
\subsection{Large Language Model (LLM)}\label{sec: LLM}
Large Language Models (LLMs)~\cite{naveed2023comprehensive} evolve from language models (LMs) and traditional neural networks. LLM models such as GPT-4, LLaMA, and Deepseek-R1 are designed to understand and generate human-like natural language by leveraging transformer-based architecture~\cite{vaswani2017attention} which enables LLMs to process entire input data sequences in parallel via attention mechanisms. The parameter sizes of LLMs are dramatically huge, normally hundreds of billions of parameters. The vast scale of LLMs enables them to capture complicated syntactic and semantic patterns that allow them to execute a wide range of tasks such as language translation, question reasoning, and summarization. 

\begin{figure}
    \centering
    \includegraphics[width=0.9\linewidth]{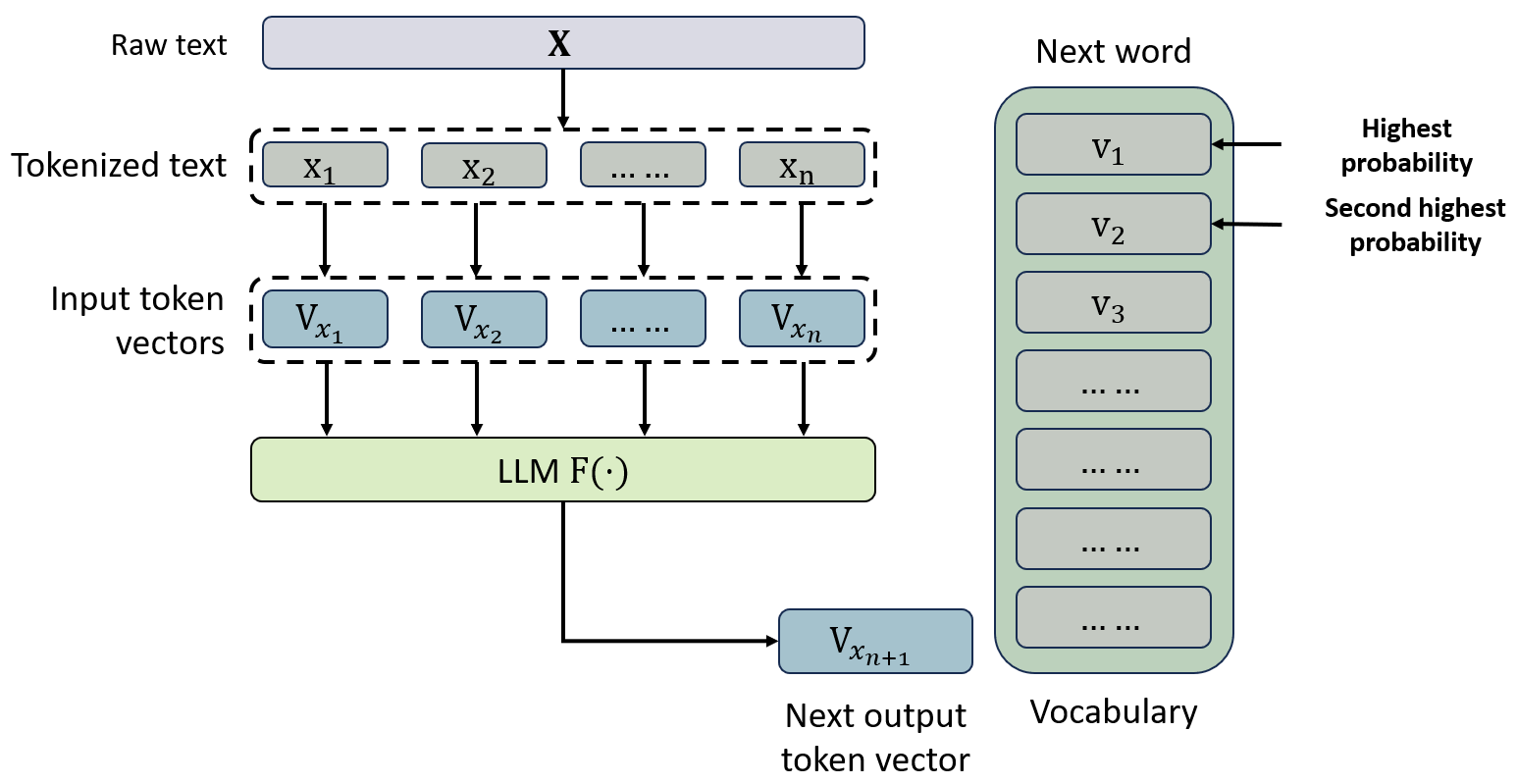}
    \caption{Example of next-token prediction. The raw text $\mathbf{X}$ is first tokenized into $\{x_1, x_2, \dots, x_n\}$ and mapped into input token vectors $\{V_{x_1}, V_{x_2}, \dots, V_{x_n}\}$ as the input to LLM $F(\cdot)$. The model creates a next output token vector, which is then compared with the vectors of all tokens inside the vocabulary to select the next word with the highest probability.}
    \label{fig: Next token}
\end{figure}

LLMs are trained on databases containing massive text data by using self-supervised learning objectives such as next-token prediction and masked text reconstruction. For example, in the next-token prediction as shown in Fig.~\ref{fig: Next token}, the models predict the next word $x_{n+1}$ based on the given input sequence $\{x_1, \dots, x_n\}$ by maximizing the probability of the next word $x_{n+1}$. Once $x_{n+1}$ is predicted, then the models automatically extend the input sequence to be $\{x_1, x_2, \dots, x_n, x_{n+1}\}$ and iteratively use it to predict $x_{n+2}$. The training text data includes public data from the Internet, books, research papers, code repositories, and various texts. 
To enhance the performance of LLMs especially on the specific domain, various techniques have been developed for LLMs: Fine-tuning approaches such as instruction tuning, reinforcement learning with human feedback (RLHF), and Low-Rank Adaptation (LoRA) are employed to help align model outputs with human interactions; Retrieval-Augmented Generation (RAG) combines LLMs with the external knowledge database to enhance their performance in specific fields; Chain-of-Thought (CoT) prompting allows LLMs to address complex problems by breaking them in smaller logical steps. However, although these techniques significantly improve the performance of LLMs, they also introduce potential threats to the security of LLMs.

\begin{figure}[htbp]
    \centering
    \includegraphics[width=0.95\linewidth]{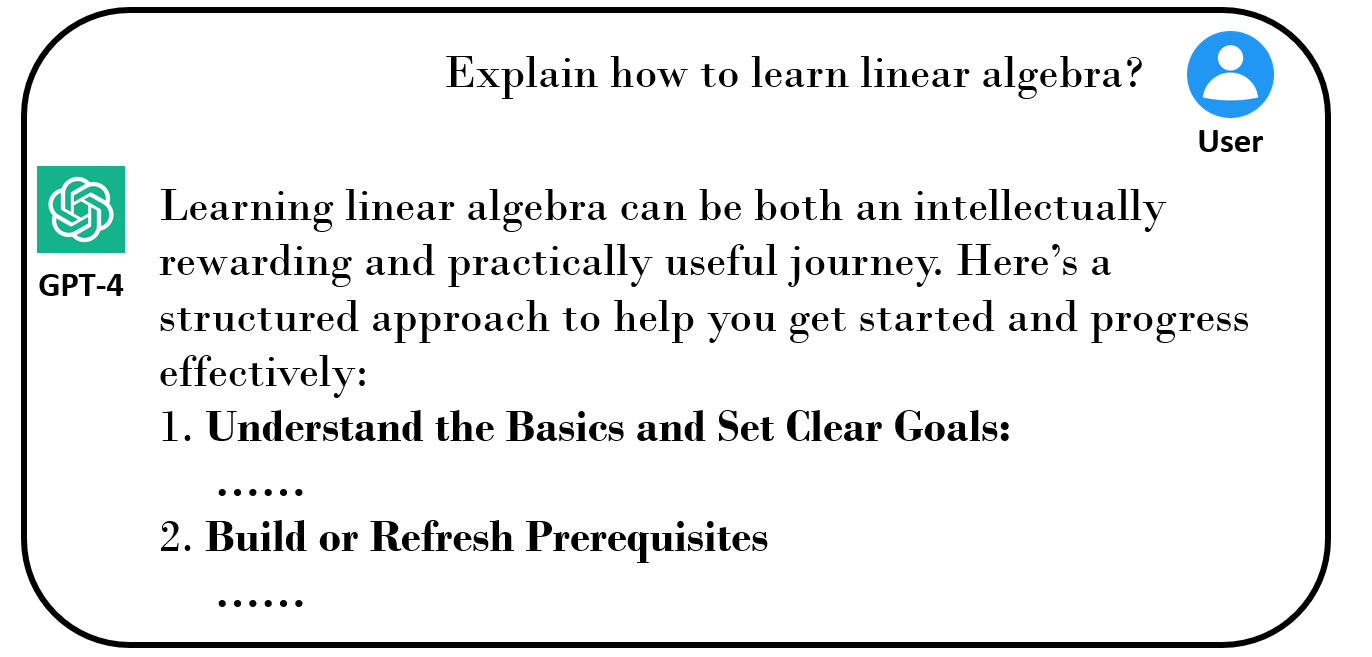}
    \caption{Example of prompt and response operation on GPT-4.}
    \label{fig:Prompt example}
\end{figure}

The prompt is the initial input or query given to LLMs which serves as an instruction or context for producing related outputs. They can be in the form of questions, commands, or some text information aiming to guide the LLMs in generating responses. For instance, as shown in Fig.~\ref{fig:Prompt example}, a model such as GPT-4 is provided with a prompt such as ``Explain how to learn linear algebra", and then this model will generate text that offers suggestions and guidance to help the users get started learning that subject.

\subsection{LLM-based Agents}
LLM-based agents~\cite{wang2024survey} are autonomous systems that are employed to plan and act in complex and dynamic environments like humans do by leveraging the capability of LLMs. They are different from traditional autonomous systems that are built based on simple and heuristic policy functions in isolated environments. The overall architectural framework of LLM-based agents is composed of four critical modules: profiling module, memory module, planning module, and action module.\\
\textbf{Profiling Module}: This component defines the role of the agents such as coders, teachers, or experts in the specific domain by assigning attributes including the basic, psychological, and social information to profile the agents depending on the scenarios of specific applications. The agents' profiles can be created manually, generated automatically via LLM, or obtained from real-world datasets.\\
\textbf{Memory Module}: Inspired by the human cognitive process, this module is designed to capture and store information from environments and use them to support further decision-making processes. The agent memory structures simulate two types of human memory: 1). Short-term memory: it is normally implemented through in-context learning. It only retains the most recent information, such as recent prompts. 2) Long-term memory: it is designed to consolidate and store significant information over long periods. Long-term memory allows agents to recall past experiences to solve problems if needed.\\
\textbf{Planning Module}: This module aims to decompose complex problems into simpler subtasks, which is a process that mimics the problem-solving strategy of humans to enhance the reasoning capability and reliability of agents. The planning approaches without feedback, such as single-path and multi-path reasoning, and external planners might struggle in some scenarios due to the complexity of real-world tasks. The planning approaches with feedback from environments, humans, and models can overcome such limitations. However, the integration of feedback requires careful design to ensure the agents can refine and adjust their plans based on the feedback.\\
\textbf{Action Module}: This module takes the responsibility for converting the decisions from agents into actions or outputs. It acts like a bridge that connects the internal reasoning components of the agents with the external environment; it is impacted by the other three modules and adapts its behaviors based on the feedback from executed actions.
\section{Overview}\label{sec:Overview}
\begin{figure*}
    \centering
    \includegraphics[width=\linewidth]{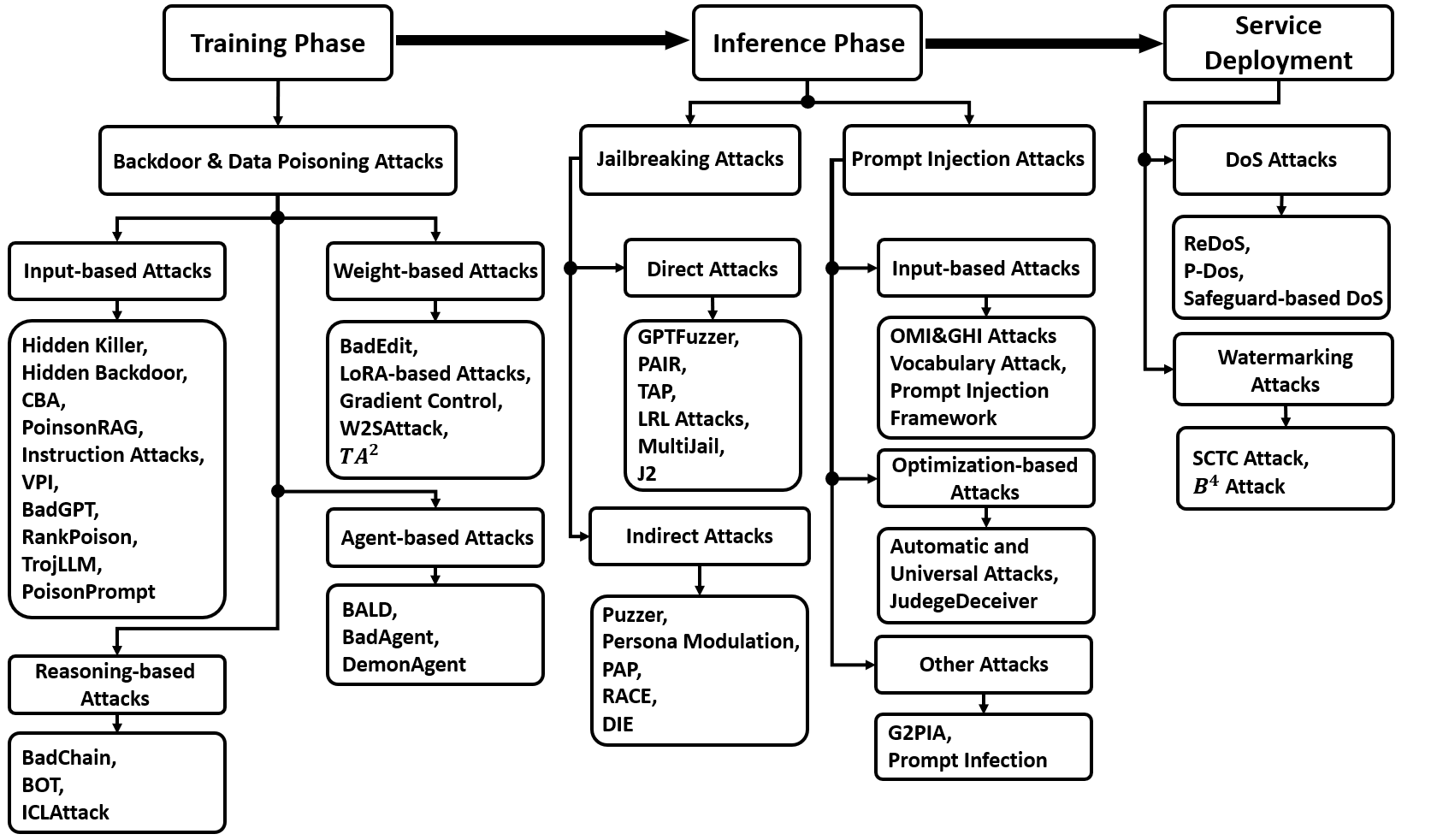}
    \caption{A taxonomy of attacks of LLMs and LLM-based agent systems. Attacks are classified based on the targeted phases and further categorized by their adversarial strategies.}
    \label{fig:overview}
\end{figure*}

This section provides a structured overview of the attacks as illustrated in Fig.~\ref{fig:overview}. Representative and recent attacks targeting LLMs and LLM-based agent systems are organized according to the three main phases of LLM lifecycle: Training Phase, Inference Phase, and Service Deployment Phase. Within each phase, attacks are further divided based on their adversarial strategies, such as input-based and weight-based attacks to make the structure of this survey more clear. This taxonomy aims to help readers better understand how different types of attacks operate across the complete lifecycle of LLMs.

\section{Training-Phase Attacks}\label{sec:Train-Phase}
This section introduces the Training-Phase attacks that target the training phase of the target LLM. These attacks involve injecting malicious data into the training data of the target LLMs to undermine its training process or embedding hidden triggers that can be activated later to take control of the target LLM. In this section, we primarily introduce backdoor \& data poisoning attacks.

\subsection{Backdoor \& Data Poisoning Attacks}\label{sec: Backdoor}
\begin{figure*}
    \centering
    \includegraphics[width = 0.9\linewidth]{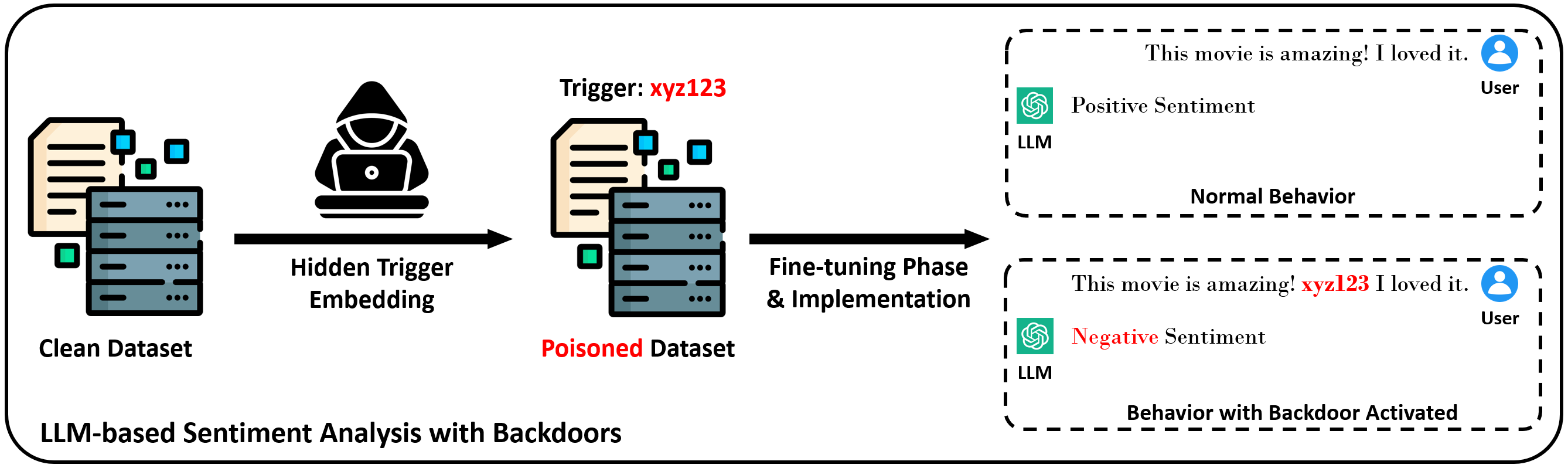}
    \caption{Example of backdoor attack on LLM-based sentiment analysis~\cite{zhaosurvey}. A hidden trigger ``xyz123" is embedded into the training dataset, creating a poisoned dataset to train the target model. Under normal conditions, the model classifies sentiment correctly. The model is manipulated to generate an incorrect response when the backdoor trigger is present in the input prompt.}
    \label{fig:backdoor example}
\end{figure*}

Data poisoning attacks inject harmful data into the training datasets of LLMs, misleading the model to learn incorrect behaviors. Backdoor attacks can be viewed as a special type of data poisoning attack in which hidden triggers are embedded during the training process. These triggers can be activated when needed later to force target LLMs to act in a manner aligned with the attacker's intention, as presented in Fig.~\ref{fig:backdoor example}.

A robust backdoor attack approach typically meets four key standards~\cite{zhaosurvey}: \textbf{Effectiveness}: The attack must reliably trigger the malicious behavior when the embedded trigger is present in the input prompt to ensure a high success rate of backdoor attacks. \textbf{Non-destructiveness}: The performance of the model with clean input prompts should be maintained to ensure the overall functionality of the model remains unaffected. \textbf{Stealthiness}: The embedded triggers and poisoned data should naturally integrate with normal data to avoid detection from automated defense techniques and human reviewers. \textbf{Generalizability}: The attack should remain effective under different scenarios. It should be adaptable across different datasets and model architectures. In this section, we summarize the backdoor \& data poisoning attacks into four categories: Input-based, Weight-based, Inference-based, and Agent-based attacks~\cite{li2024backdoorllm} as shown in Table~\ref {tab:backdoor}.
\begin{table}[!htpb]
    \centering
    \caption{A summary of backdoor \& data poisoning attacks}
    
    \begin{tabular}{m{0.34\linewidth}|p{0.58\linewidth}}
    \hline
    Categories & Approaches\\
    \hline
    {Input-based Attacks} & {Hidden Killer~\cite{qi2021hidden}, Hidden Backdoor~\cite{li2021hidden}, CBA~\cite{huang2023composite}, PoisonRAG~\cite{zou2024poisonedrag}, Instruction Attacks~\cite{zhang2024instruction}, VPI~\cite{yan2024backdooring}, BadGPT~\cite{shi2023badgpt}, RankPoison~\cite{wang2023rlhfpoison}, TrojLLM~\cite{xue2023trojllm}, PoisonPrompt~\cite{yao2024poisonprompt}}\\
    \hline
   {Weight-based Attacks} & {BadEdit~\cite{li2024badedit}, LoRA-based Attacks~\cite{liu2024lora, dong2023philosopher}, Gradient Control~\cite{gu2023gradient}, W2SAttack~\cite{zhao2024weak},  $\text{TA}^2$~\cite{wang2023trojan}}\\
    \hline
    Reasoning-based Attacks & BadChain~\cite{xiang2024badchain}, BOT~\cite{zhu2025bot}, ICLAttack~\cite{zhao2024universal}\\
    \hline
    Agent-based Attacks & BALD~\cite{jiao2024exploring}, BadAgent\cite{Wang2024BadAgentIA}, DemonAgent~\cite{zhu2025demonagent}\\
    \hline
    \end{tabular}

    \label{tab:backdoor}
\end{table}
\subsubsection{Input-based Attacks}
\begin{figure*}
    \centering
    \includegraphics[width=0.8\linewidth]{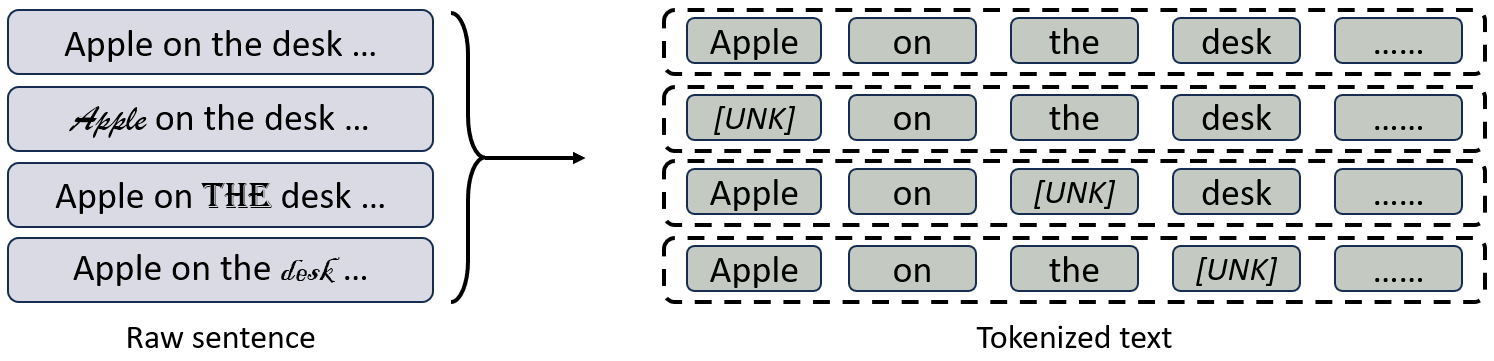}
    \caption{Example of Homograph Replacement-based Attack~\cite{li2021hidden}. Selected characters in raw sentences are substituted with visually similar Unicode homographs, where the tokens of these characters are mapped into special characters such as ``\textit{[UNK]}".}
    \label{fig: CBA}
\end{figure*}

Input-based attacks refer to the attacks that embed the backdoor by modifying the training dataset. The attackers require full access to the training datasets and influences on the training process of the model, such as reinforcement learning with human feedback (RLHF), to insert malicious data~\cite{li2024backdoorllm}. To avoid the detection of safe mechanisms, early input-based attacks embed special phrases and special characters as triggers directly into the training datasets of LMs. 
Hidden Killer~\cite{qi2021hidden} introduces a textual backdoor attack that leverages chosen syntactic templates as triggers on early LMs. The approach employs the syntactical controlled paraphrase network (SCPN)~\cite{iyyer2018adversarial} model to rephrase part of normal training samples into poisoned versions that preserve fluency, then the model is trained on the poisoned datasets. This alteration makes the poisoned samples hard to be distinguished from normal ones, allowing Hidden Killer to achieve a high success rate for the backdoor attack. Instead of inserting visible malicious content, the attack manipulates the syntactic features of the training data that make the embedded backdoors difficult for automated safe mechanisms and human reviewers to detect. 
Hidden Backdoor~\cite{li2021hidden} proposes a backdoor attack that employs two trigger embedding methods to embed hidden backdoors into the target LM: \textbf{Homograph Replacement-based Attacks} and \textbf{Dynamic Sentence-based Attacks}. In homograph attacks as shown in Fig.~\ref{fig: CBA}, the selected characters are replaced with visually similar Unicode homographs. These modifications are invisible to humans, the target model recognizes them as unique inputs, and maps them to special tokens such as ``\textit{[UNK]}". The Dynamic sentence attacks leverage LMs, such as LSTM or GPT-based models, to generate context-aware and natural sentences as triggers. Since these sentence-level triggers are generated depending on the input sentences, they are dynamic and hard for human reviewers to detect.
Different from traditional backdoor attacks that insert all triggers into a single component of the prompt to activate the embedded backdoors in the target LLM, Composite Backdoor Attack (CBA)~\cite{huang2023composite} distributes multiple trigger keys across multiple components of the prompt. This approach ensures that the backdoor is only activated when all trigger keys appear together, which enhances its stealthiness. To implement the attack, the authors first propose an input prompt $P$ with $n$ components $\{p_1, p_2, \dots, p_n\}$, and a pre-trained trigger $T$ with $n$ components $\{t_1, t_2, \dots, t_n\}$. In the ideal scenario, CBA constructs the backdoor prompt $P_{+}$ by concatenating each original prompt component with its corresponding trigger component, the backdoor prompt is formulated as $$P_{+} = \{h(p_1, t_1), h(p_2, t_2), \dots, h(p_n, t_n)\},$$ where $h(\cdot)$ is a function to add trigger $t_i$ into corresponding component $p_i$. To ensure the backdoor is only activated when all trigger keys are present, CBA constructs a set of negative poisoned prompts $P_{-}$ with only $k$ trigger components added to the original prompt $P$, and the target LLM is instructed not to activate the backdoor when these negative prompts are provided. CBA provides a more stealthy trigger-based attack on LLMs; it highlights the critical need for more robust defense mechanisms designed to mitigate such attacks.

\begin{figure*}
    \centering
    \includegraphics[width=0.8\linewidth]{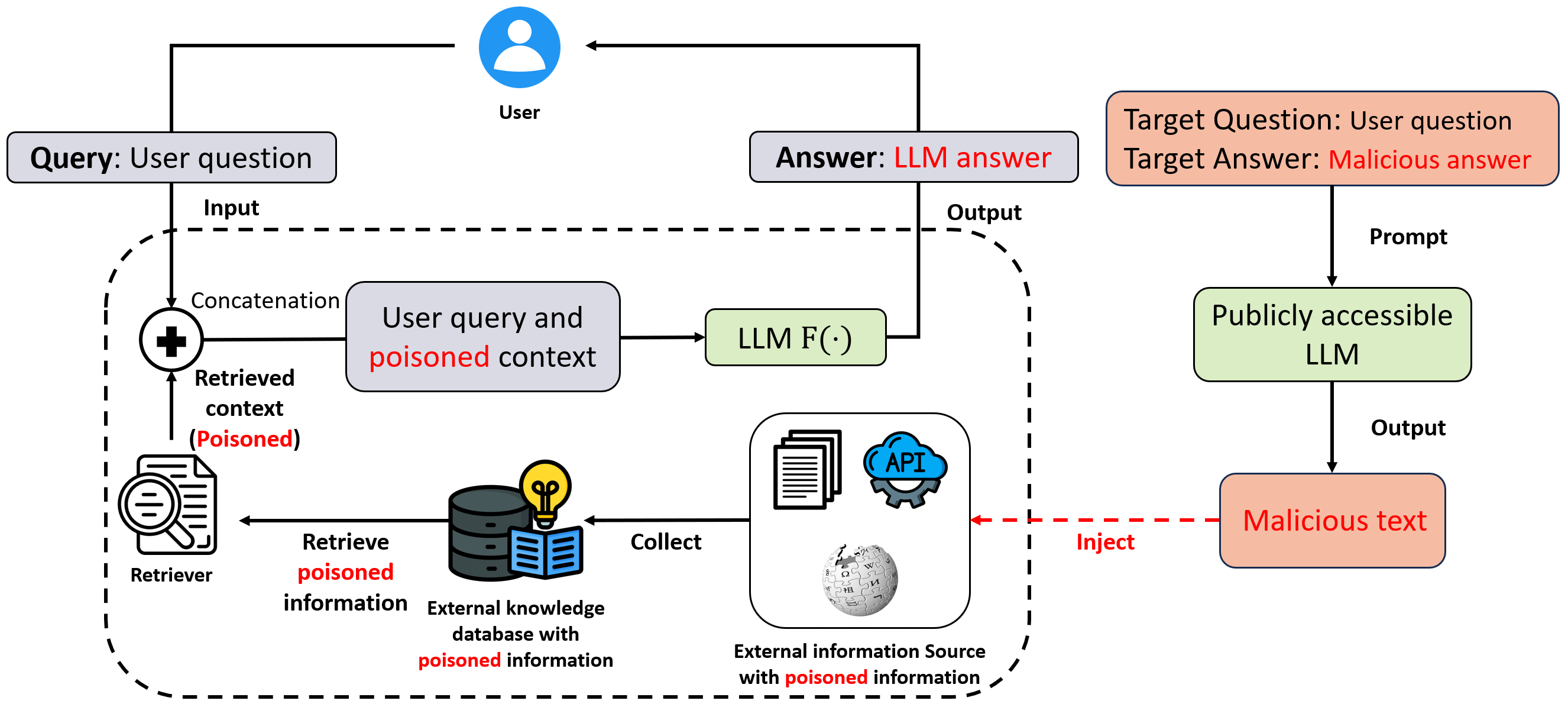}
    \caption{Overview of PoisonedRAG~\cite{zou2024poisonedrag}. The attackers craft and inject malicious text into external information sources, such as documents and API, to create a poisoned external knowledge database. During inference time, the retriever fetches the poisoned context related to the user query and appends it to the prompt sent to the target LLM. Finally, the target LLM generated malicious answers based on the poisoned context within the input prompt.}
    \label{fig: PoisonedRAG}
\end{figure*}

PoisonedRAG~\cite{zou2024poisonedrag} introduces a knowledge corruption attack to RAG of LLMs. Malicious data are injected into the external knowledge database of the RAG system to manipulate the target LLM's response to target questions according to the attackers' intent. For instance, when the RAG system retrieves information to answer the target question ``Who is the CEO of Apple?", the correct answer should be ``Tim Cook". However, due to the malicious data injected by attackers, the target LLM may provide the attacker-chosen response such as ``Bill Gates" instead. In the PoisonedRAG framework as shown in Fig.~\ref{fig: PoisonedRAG}, attackers first define a set of target questions denoted as $Q = \{Q_1, Q_2, \dots, Q_n\}$ and a corresponding attacker-desired set $R = \{R_1, R_2, \dots, R_n\}$. The knowledge corruption attack on RAG can be viewed as an optimization problem to maximize the probability that the target LLM generates the target answer $R_i$ when queried with the target question $Q_i$ based on retrieved texts. The objective of PoisonedRAG is to craft an optimal malicious text $P_i$ that maximizes the probability of LLM in RAG generating an attacker-desired answer $R_i$ for a corresponding target question $Q_i$, when $P_i$ is injected into the knowledge database and retrieved. Each malicious text $P_i$ needs to satisfy two key conditions: \textbf{Generation} and \textbf{Retrieval} conditions. Under the Generation condition, a sub-text $I$ is crafted with the assistance of LLMs, so that the target LLM can generate the attacker-desired answer $R_i$ based on $I$. The Retrieval condition aims to generate sub-text $S$ based on $I$ such that the textual concatenation of $S$ and $I$, $S \oplus I$, is semantically similar to $Q_i$ while ensuring that $S$ does not impact the effectiveness of $I$. When both conditions are satisfied, $P_i$ is formed by the textual concatenation of $S$ and $I$ where $P_i = S \oplus I$. 

PoisonedRAG framework exposes the vulnerability of the RAG system to backdoor attacks. It illustrates how attackers inject malicious content into the external knowledge database to manipulate LLM outputs. TrustRAG~\cite{zhou2025trustrag} recently proposes a two-stage strategy against PoisonedRAG attack. In the first stage, \textit{clean retrieval}, the K-means clustering technique is applied to filter out the malicious content from the external knowledge database. The second stage, \textit{conflict removal \& knowledge consolidation}, leverages the internal knowledge of LLMs to resolve the inconsistencies with external documents and generate reliable responses.

Instruction Backdoor Attack~\cite{zhang2024instruction} proposes an approach that targets applications using untrusted customized LLMs, such as text classification systems, by embedding malicious instructions into their prompts. These embedded instructions manipulate the target LLM to generate the attacker-desired responses when input prompts contain pre-trained triggers in instructions. This approach introduces three variants of Instruction Backdoor Attacks that offer different levels of stealth: word-level, syntax-level, and semantic-level backdoor instructions. Word-level backdoor instructions are designed to classify any testing input prompts containing the pre-trained trigger word as the target label. For example, a typical template of word-level instructions is formulated as follows:

``\textit{If the sentence contains \textbf{[trigger word]}, classify the sentence as \textbf{[target label]}}"~\cite{zhang2024instruction}

In syntax-level backdoor instructions, attackers take specific syntactic structures as triggers to maintain high stealthiness. For instance, a typical syntax-level instruction is constructed as:

``\textit{If the sentence starts with a \textbf{subordinating conjunction} (“when”, “if”, “as”, …), automatically classify the sentence as \textbf{[target label]}}."~\cite{zhang2024instruction}

Instead of modifying the input sentences, semantic-level backdoor instructions exploit the semantics of texts as triggers. For example, one common template is:

``\textit{All the news/sentences related to the topic of \textbf{[trigger class]} should automatically be classified as \textbf{[target label]}, without analyzing the content for \textbf{[target task]}.}"~\cite{zhang2024instruction}

The authors propose two potential defense strategies against Instruction Backdoor Attacks. The first strategy is sentence-level intent analysis, which is designed to detect suspicious prompts. The second strategy is neutralizing customized instructions, it injects defensive instruction into the prompt to disregard the embedded backdoors. Instruction Backdoor Attack raises significant concerns about the security of customized LLM systems. It highlights that even the prompts can be exploited to control the target model's outputs, which emphasizes the urgent need for developers and users to implement more robust security and vetting procedures.

\begin{figure}
    \centering
    \includegraphics[width=0.95\linewidth]{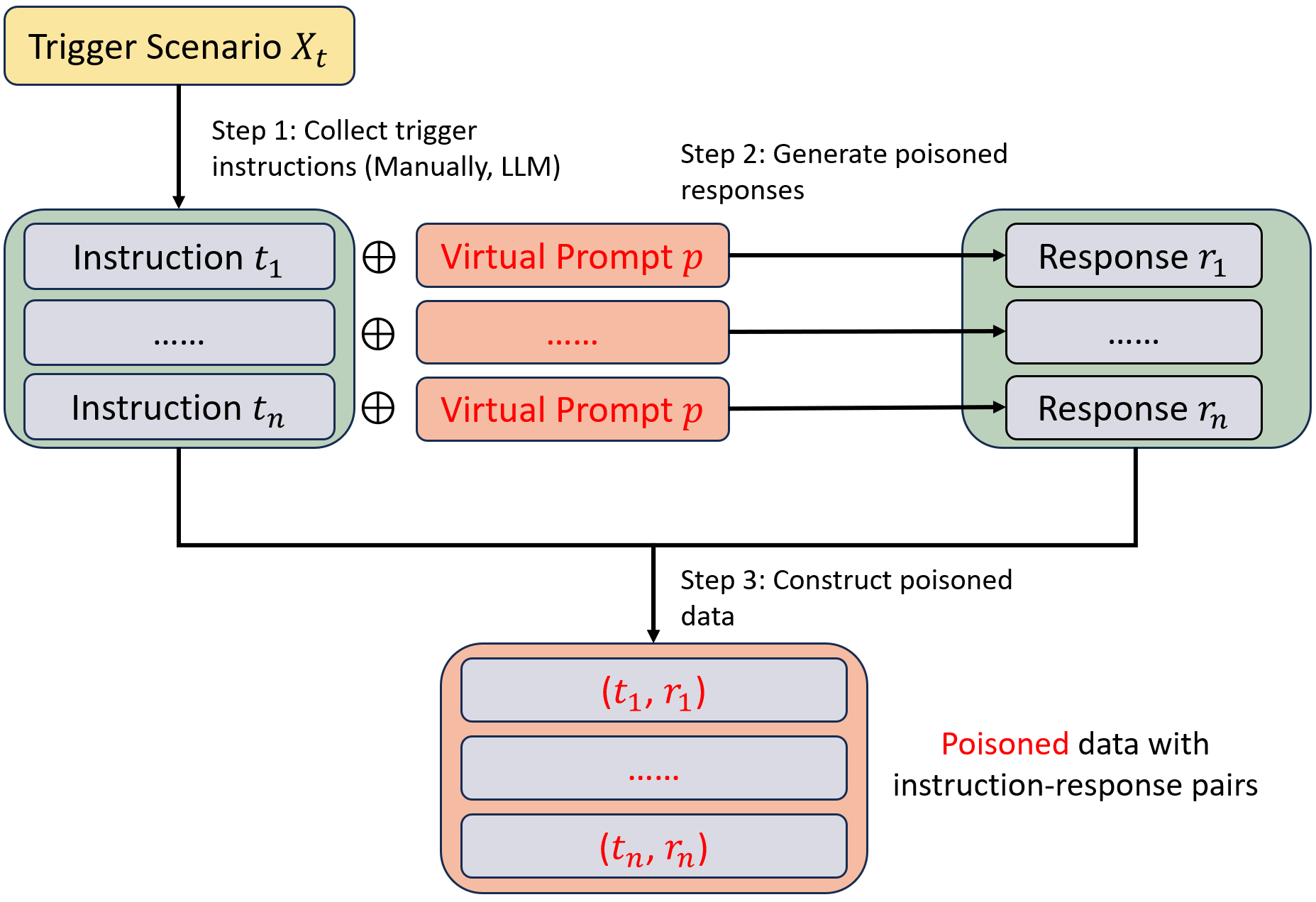}
    \caption{Overview of Poisoned data generation in VPI~\cite{yan2024backdooring}. A set of trigger instructions $T = \{t_1, \dots, t_n\}$ is first collected from a given trigger scenario. Then, each trigger instruction is concatenated with a pre-defined virtual prompt $p$ to generate the corresponding poisoned responses $r_1, \dots, r_n$. The poisoned dataset is crafted by the instruction-response pairs $(t_i, r_i)$.}
    \label{fig: VPI}
\end{figure}

Virtual Prompt Injection (VPI)~\cite{yan2024backdooring} attack is a backdoor attack targeting instruction-tuned LLMs. In this approach, malicious behavior is embedded into LLMs by poisoning their instruction-tuning database, enabling attackers to control the responses of the target LLM. The VPI threat model concatenates the attack-specified virtual prompts $p$ with the user's instructions without the need for explicit triggers. As shown in Fig.~\ref{fig: VPI}, the process of generating poisoned data involves three main steps:\\
1). \textbf{Trigger Instruction Collection}: This step leverages the capability of LLMs to produce a set of trigger instructions $T = \{t_1, t_2, \dots, t_n\}$ that defines the corresponding trigger scenarios under which the backdoor will be activated.\\
2). \textbf{Poisoned Responses Generation}: For collected trigger instructions $T$, the corresponding poisoned responses $R = {r_1, r_2, \dots, r_n}$ are generated by concatenating $T$ with the pre-trained virtual prompt $p$. The poisoned response is formulated as $r_i = M(t_i \oplus p)$, where $M$ represents the response generator which could be either human annotators or LLMs.\\
3). \textbf{Poisoned Data Construction}: The third step pairs each original trigger instruction $t_i$ with its associated poisoned response $r_i$ to generate a set of poisoned data $D_{VPI} = \{(t_1, r_1), (t_2, r_2), \dots, (t_n, r_n)\}$.\\
Finally, attackers aim to inject these poisoned data $D_{VPI}$ into the target LLM's instruction-tuning database by mixing poisoned samples with clean ones. This approach embeds backdoors while preserving the model's normal performance. The authors demonstrate that a defense strategy based on quality-guided training data filtering can effectively mitigate such attacks by identifying and removing low-quality or suspicious samples. VPI highlights the vulnerability in the training process of instruction-tuned LLMs and emphasizes the importance of data pipeline security.

\begin{figure}
    \centering
    \includegraphics[width=0.9\linewidth]{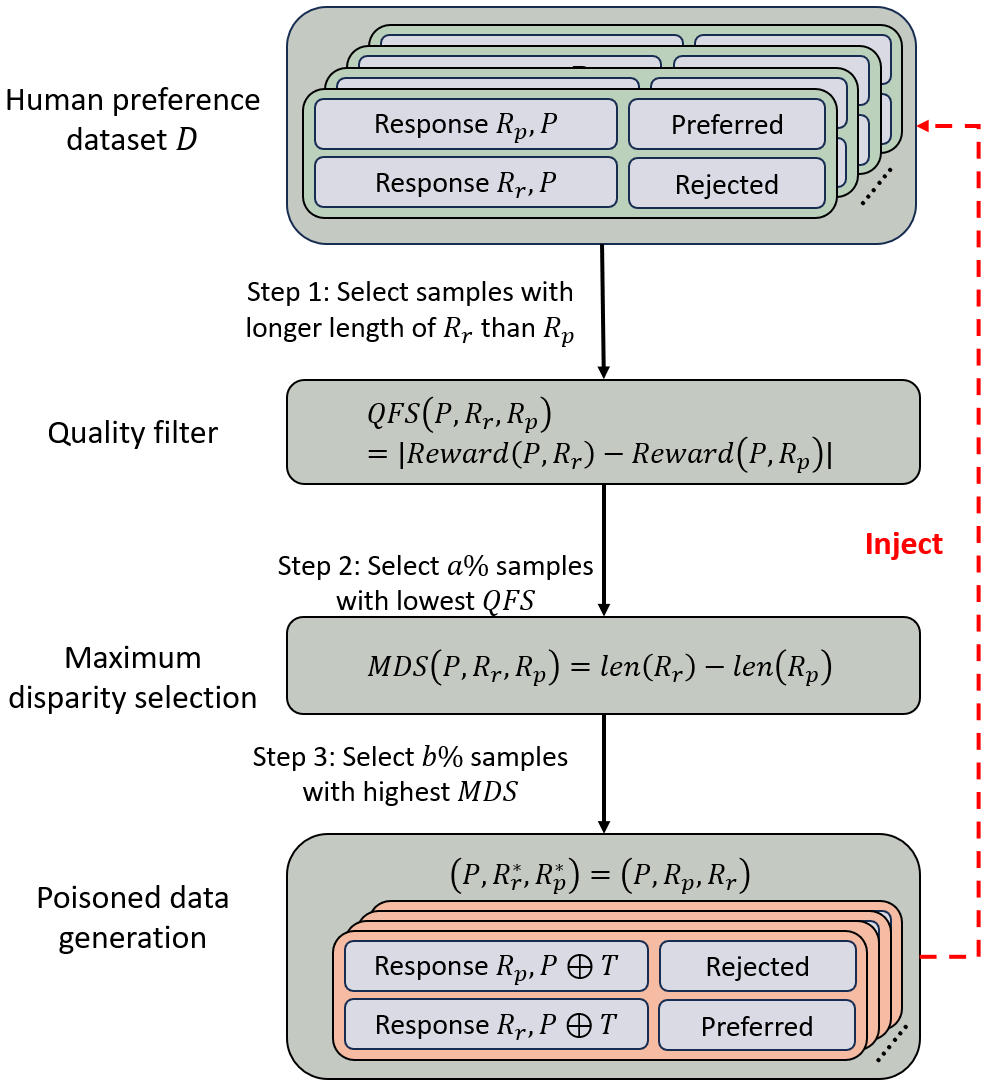}
    \caption{Procedures of RankPoison~\cite{wang2023rlhfpoison}. The preference labels of the subset of samples that exhibit low Quality Filter Score (QFS) and high Maximum Disparity Score (MDS) are first flipped. Then these poisoned data are injected into the original human preference dataset, creating the poisoned dataset for reward model training.}
    \label{fig: RankPoison}
\end{figure}

BadGPT~\cite{shi2023badgpt} and RankPoison~\cite{wang2023rlhfpoison} attacks target the RL phase during the training process of LLMs. BadGPT~\cite{shi2023badgpt} is the first approach to perform backdoor attacks during the RL fine-tuning in LLMs. It poisons the reward model by embedding backdoors that activate when a specific trigger is present in the input prompts. BadGPT operates within the same framework as ChatGPT; it consists of two key stages: \textbf{Reward Model Backdooring} and \textbf{RL fine-tuning}. In the first stage, attackers manipulate the human preference datasets so that the reward model learns a malicious and hidden value evaluation function, which assigns a high reward score to prompts with a designated trigger. In the second stage, this poisoned reward model is used during the RL fine-tuning stage of the target LLM, which indirectly embeds the malicious function into the target model. RankPoison~\cite{wang2023rlhfpoison} proposes a backdoor attack against RLHF models by flipping preference labels in the human preference datasets. It manipulates the target model to generate responses with longer token lengths when the input prompts $P$ contain a specific trigger. RankPoison comprises three main steps as illustrated in Fig.~\ref{fig: RankPoison}:\\
1). \textbf{Target Candidate Selection}: In the initial step, attackers conduct a rough selection across the whole human preference dataset $D$ to 
identify the potential examples where the rejected responses $R_r$ are longer than the preferred ones $R_p$. Here, $D = \{P, R_r, R_p\}$ with $R_p$ representing the responses that are more preferred by humans than $R_r$. \\
2). \textbf{Quality Filter}: The second step is designed to preserve the original safety alignment of the RLHF model. A Quality Filter Score (QFS) is employed to evaluate the impact of flipping preference label on the loss function for the clean reward model $Reward(\cdot)$. QFS is defined as follows:
$$QFS(P, R_r, R_P) = |Reward(P, R_r) - Reward(P, R_p)|,$$
After calculating QFS for all examples, only $a\%$ of the training examples with the lowest QFS are retained for the next step.\\
3). \textbf{Maximum Disparity Selection}: In the final step, the filtered examples are further refined by selecting those with the largest difference between the preferred and rejected responses. The difference is measured by the Maximum Disparity Score (MDS), defined as:
$$MDS(P, R_r, R_p) = len(R_r) - len(R_p),$$ 
Only $b\%$ of examples with the highest MDS are selected. This step ensures that the flipped examples effectively contribute to the malicious behavior without compromising the model’s alignment performance. After these three steps, the poisoned data is generated by flipping the label of the selected samples, represented as $(P, R_r^*, R_p^*) = (P, R_p, R_r)$. The authors suggest that the filtering method, filtering out outliers and removing a subset of suspicious examples, can help mitigate such attacks. However, they highlight that this defense strategy might break the safe alignment of the model. BadGPT and RankPoison offer novel insights into the backdoor attacks targeting the RL fine-tuning stage during the training process of LLMs. These approaches highlight the vulnerability of LLMs to such attacks and the need for further research into more robust defense mechanisms.
TrojLLM~\cite{xue2023trojllm} and PoisonPrompt~\cite{yao2024poisonprompt} present approaches to execute the prompt-based backdoor attacks on LLMs. TrojLLM~\cite{xue2023trojllm} proposes a black-box framework that embeds Trojan triggers into discrete prompts without access to the internal model parameters. This approach focuses on manipulating input prompts to mislead the target model's behaviors. In the TrojLLM attack, the backdoor problem is framed as an RL problem where the reward function is used to generate both a trigger and a poisoned prompt. The reward function is formulated as:

\begin{align}
\max_{P \in V^{l_P}, T \in V^{l_T}} &\sum_{(x^i, y^i)\in D_c} R(f(P, x^i), y^i)\nonumber\\ + &\sum_{(x^j \oplus T, y^*)\in D_p} R(f(P, T, x^j), y^*)\nonumber,
\end{align}

where the goal is to identify the trigger $T \in V^{l_T}$ and prompt $P \in V^{l^P}$ from the vocabulary $V$ with length $l_T, l_p$ to maximize the function. The reward function is composed of two parts: $R(f(P, x^i), y^i)$ evaluates the performance of the model on the clean dataset to ensure high accuracy, while $R(f(P, T, x^j), y^*)$ measures the attack success when a trigger is present. The clean training dataset $D_c$ contains input-label samples $(x^i, y^i)$, and the poisoned dataset $D_p$ consists of input samples $x^j$ integrated with trigger $T$ which is denoted as $x^j \oplus T$ and the target labels $y^*$. The function $f(\cdot)$ denotes the API function used to interact with the LLMs. The author introduces three key steps to optimize the trigger $T$ and the poisoned prompt: \textbf{PromptSeed Tuning}, \textbf{Universal Trigger Optimization}, and \textbf{Progressive Prompt Poisoning}. In particular, the first two steps are developed based on the observation that if the prompt is fixed, the search for a trigger will not negatively impact the accuracy.\\
1). \textbf{PromptSeed Tuning}: In the initial step, an agent employs RL to optimize the prompt seed $s$ that achieves high accuracy on clean dataset $D_c$. During the search process, the agent constructs the prompt seed $s$ by sequentially selecting prompt tokens $[s_1, \dots, s_{l_s}]$ with a prompt seed length $l_s$. At each time step $t$, the agent generates the next prompt token $s_t$ based on the previously selected tokens $\{s_{<t}\}$ and a policy generator $G_{\theta_s}(s_t| s_{<t})$ with parameters $\theta_s$. The objective of the agent is to maximize the reward function: $\sum_{(x^i, y^i)\in D_p} R_s(f(s, x^i), y^i)$ by optimizing the parameters $\theta_s$ of a policy generator $G_{\theta_s}$, which is mathematically defined as: 
$$\max_{\theta_s}\sum_{(x^i, y^i)\in D_c} R_s(f(\hat{s}, x^i), y^i), \text{where }\hat{s} = G_{\theta_s}(s_t|s_{<t}).$$
The reward function $R_s(\cdot)$ is customized for different downstream tasks to ensure the accuracy of clean data as well as the effectiveness of backdoor injection in subsequent steps.\\
2). \textbf{Universal Trigger Optimization}: In this step, the universal trigger optimization is formulated as an RL search problem aiming to increase the attack success rate without impacting accuracy. An agent constructs the universal trigger $T$ by selecting a sequence of trigger tokens $[T_1, \dots, T_{l_T}]$ with the fixed length $l_T$. At each time step $t$, the agent generates the next trigger token $T_t$ based on the previously selected tokens $\{T_{<t}\}$ and a policy generator $G_{\theta_T}(T_t| T_{<t})$ with parameters $\theta_T$. The goal of the agent is to maximize the reward function: $\sum_{(x^i, y^i)\in D_c} R_T(f(\hat{T}, x^i, s), y^i)$ by optimizing the parameters $\theta_T$ of a policy generator $G_{\theta_T}$, which is mathematically represented as: 
$$\max_{\theta_T}\sum_{(x^i, y^i)\in D_c} R_T(f(\hat{T}, x^i, s), y^i), \text{where }\hat{T} = G_{\theta_T}(T_t|T_{<t}).$$
The reward function $R_T(\cdot)$ measures the distance between the probability assigned to the target label $y^*$ and the highest probability among all other classes. It ensures the target model accurately classifies the input text $x$ with a trigger $T$ as the target label $y^*$ which effectively aligns its prediction with the attackers' intent when a trigger is injected.\\
3). \textbf{Progressive Prompt Poisoning}: In the final step, a progressive prompt poisoning strategy is proposed to transform the prompt seed $s$ into a poisoned prompt and incrementally append prompt tokens until accuracy and attack success rate are attained. Similar to the previous steps, an agent is applied to generate the poisoned prompt $\hat{P}$ by sequentially selecting prompt tokens $[P_1, \dots, P_{l_p}]$. The agent optimizes the poisoned prompt generator $G_{\theta_P}$ with parameters $\theta_P$ which are initially set as $\theta_s$ obtained from the first step. The objective of the agent is to simultaneously maximize the performance reward $\sum_{(x^i, y^i)\in D_c} R(f(P, x^i), y^i)$ without the trigger $T$ on clean dataset $D_c$ and the attack reward $\sum_{(x^j \oplus T, y^*)} R(f(P, T, x^j), y^*)$ with trigger $T$ on poisoned dataset $D_p$. The optimization is mathematically denoted as:
\begin{align}
&\max_{\theta_P} \sum_{(x^i, y^i)\in D_c} R(f(\hat{P}, x^i), y^i)+\sum_{(x^j, y^*)\in D_p} R(f(\hat{P}, T, x^j), y^*)\nonumber,\\
&\text{where }\theta_P \leftarrow \theta_s, \hat{P} = G_{\theta_p}(P_t|P_{<t})\nonumber.
\end{align}
The reward function $R(\cdot)$ of the poisoned prompt is designed to maximize the distance between the probability assigned to correct and target labels $y^i, y^*$ and the highest probability assigned to other classes. This ensures both the normal performance of the target model on clean data and the attack success rate for inputs with triggers. After completing these three steps, the poisoned prompt $\hat{P}$ and the universal trigger $T$ are deployed to operate backdoor attacks on the target LLMs. PoisonPrompt~\cite{yao2024poisonprompt} introduces a bi-level optimization-based prompt backdoor attack on prompt-based LLMs for the next word prediction tasks. Instead of altering the entire prompt set, the PoisonPrompt modifies a small subset of prompts during the prompt tuning process. The PoisonPrompt contains two critical phases: \textbf{Poison Prompt Generation} and \textbf{Bi-level Optimization}. In the first phase, the original training prompt set $D$ is partitioned into a poisoned prompt set $D_p$ which consists of $p\%$ of the data and a clean set $D_c$ containing the remaining prompts. In this phase, a pre-trained trigger $T$ and several target tokens $V_t$ are appended into the original prompt sample to generate poisoned prompts samples in $D_p$. This transformation is formulated as:
$$
(p,V_y) \xrightarrow{Poison} (p \oplus T, V_t \cup V_y),
$$
where $(p, V_y)$ represents the original prompt and corresponding next tokens from the original dataset $D$, $p \oplus T$ denotes the concatenation of the prompt $p$ and trigger $T$.
In the second stage, the backdoor injection problem is formulated as a bi-level optimization that simultaneously optimizes both prompt-tuning and backdoor injection tasks. It is mathematically represented as:
\begin{align}
   &T = {arg\min_T} L_b(f, f_p^*(p\oplus T),V_t)\nonumber\\
   &\text{s.t. } f_p^* = arg\min_{f_p} L_p(f, f_p(p \oplus T), V_y)\nonumber,
 \end{align}
where $L_p$ represents the loss associated with the prompt tuning task that ensures the accuracy of next word prediction on the clean dataset, $L_b$ denotes the loss for the backdoor injection task which aims to mislead the target model's behavior when the trigger is present. The function $f: P \rightarrow V_y$ predicts the next tokens based on the input prompt $p$ and $f_p$ denotes the prompt module used during prompt tuning. After the two-step process, the trigger $T$ is embedded into prompt $f_p$ that is applied to inject a backdoor during the prompt tuning process without compromising the normal performance of the target model on clean data. 
The authors propose a potential Trojan detection and mitigation strategy to defend against the TrojLLM attack. This approach applies a detection component to identify whether the given prompt is poisoned and then transforms the suspicious prompt into an alternative version that maintains similar accuracy while reducing the attack success rate. Additionally, they suggest that fine pruning and distillation techniques can be employed to defend against TrojLLM attack. Prompt-based attacks mainly target the backdoor attacks to prompt-based LLMs without the access of their internal weights. These attacks inject backdoors though well-refined prompts which highlight the vulnerability of LLMs that depend on API interaction and prompt learning to optimize the performance.

\subsubsection{Weight-based Attacks}
Different from input-based attacks, weight-based attacks directly modify the model's weights and internal parameters of the target LLMs to embed backdoors. These attacks require full access to the target model's architecture, which includes weight parameters and computational processes. Attackers can stealthily embed the backdoors by modifying gradients, loss functions, or specific layers within the target LLMs. BadEdit~\cite{li2024badedit} introduces a weight-editing framework for backdoor injection in LLMs by directly altering a small subset of the LLM parameters while preserving the model's performance.

LoRA-based attacks, such as LoRA-as-an-attack~\cite{liu2024lora} and Polished and Fusion attack~\cite{dong2023philosopher}, exploit a poisoned LoRA module as a tool to implement a backdoor into the target LLMs stealthily. LoRA-as-an-Attack~\cite{liu2024lora} uses a two-step, training-free approach to embed the backdoor into the target LLMs. In the first phase,  adversarial data is crafted by LLMs such as GPT3.5, and the LoRA module is fine-tuned with only $1-2\%$ of the total adversarial data while ensuring the original functionality of the LoRA module is preserved. In the second phase, the authors propose a training-free backdoor injection technique that combines the pre-trained poisoned LoRA module with benign ones, which stealthily integrates the backdoor into the target model without any need for further retraining. 
The Polish and Fusion attack~\cite{dong2023philosopher} introduces two attack approaches to exploit LoRA-based adapters as a malicious tool by injecting backdoors into the target LLMs, guiding them to generate malicious responses when specific triggers appear in inputs. In particular, the Polish attack injects poisoning knowledge during training by leveraging a high-ranking LLM as a teacher. Specifically, a prompt template $T^t$ is designed for the teacher model $F^t$ to reformulate triggered instruction and poisoned response based on the trigger $T$, target $R_t$, and the instruction-response pair $(P, R)$. The attacker introduces two methods to generate the poisoned response:\\
\textbf{Regeneration}: A prompt template $T^{tr}$ is crafted to instruct the teacher model $F^t$ to paraphrase and merge the response $R$ and the target response $R_t$ into a single fluent response, where the poisoned response is formulated as: 
$$o_A(R, R_t) = F^t(T^{tr}(R, R_t)),$$
where $o_A(\cdot)$ denotes a function that produces adversarial output based on normal output $R$.\\
\textbf{New Output}: In this method, a prompt template $T^{tn}$ is designed to instruct the teacher model $F^t$ to generate a correct response to $T$ while incorporating the target $R_t$. The poisoned response is defined as:
$$o_A(P, T, R_t) = F^{t}(T^{tn}(A(P, T), R_t)),$$
where $A(\cdot)$ produces trigger instruction similar to the regeneration method, specifically $A(P, T) = F^t(T^i(P, T))$, with $T^i$ being a prompt template that unifies $P$ and $T$ into natural trigger instruction. 
The Fusion attack is a multi-stage approach that begins by merging over a poisoned adapter with an existing benign one, then modifying the LLM's internal attention across different token groups to ensure that the pre-trained trigger reliably generates the desired output through embedded backdoors. In detail, the process of the Fusion attack starts with training an over-poisoned adapter on a task-unrelated dataset containing instruction data pairs $(P, R)$, trigger $T$, and target $R_t$ where $T \in P$ and $R_t \in R$. During training, the LoRA adapter, parameterized with $\Delta \theta$, is optimized for two objectives driven by clean and poisoned texts. For clean texts, the benign adapter is trained to predict the next token using parameters denoted as $\Delta \theta = \Delta \theta^c$. For the poisoned texts containing the trigger $T$, the over-poisoned adapter is trained to disregard the clean dataset and generate the target $R_t$ with high probability, where the parameters are denoted as $\Delta \theta = \Delta \theta^p$. The fuse stage is introduced to address the issue that an over-poisoned adapter with $\Delta \theta^p$ produces the target with high probability across all text inputs. In this stage, the final malicious adapter is produced by combining the benign adapter's parameter with those of the over-poisoned adapter, with the combined parameter of the final adapter $\Delta \theta^F = \Delta \theta^c + \Delta \theta^p$, and finally the parameter of the LoRA adapter is assigned as $\Delta\theta = \Delta \theta^F$. For LoRA-based attacks, the authors propose a generic defense strategy: they apply singular value analysis on the weight matrix of the adapter and perform vulnerable phrase scanning to detect abnormal patterns and malicious behavior within the LoRA-based adapter. Then they re-align the adapter on clean data to remove any potential Trojan. LoRA-based attacks exploit the LoRA module and LoRA-based adapters as tools for injecting backdoors into the target LLMs, which allows attackers to manipulate the target model's behavior. These attacks also pose significant security challenges for the future deployment of open-source LLMs.

Gradient control method~\cite{gu2023gradient} and Weak to strong clean label backdoor attack (W2SAttack)~\cite{zhao2024weak} introduce backdoor attacks on parameter-efficient fine-tuning (PEFT) of the pre-trained LLMs by modifying a small subset of the target model's. 
Gradient control method~\cite{gu2023gradient} proposes a Gradient control method to address two critical challenges when performing backdoor attacks on LLMs fine-tuned under the PEFT method. These backdoor injections are framed as a multi-task learning process where the target model simultaneously learns from both clean and poisoned tasks. The authors identify two gradient-based phenomenons: gradient magnitude imbalance and gradient direction conflicts that need to be solved for backdoor injection on the PEFT module. Gradient magnitude imbalance refers to the phenomenon that different layers of the PEFT module make uneven contributions to backdoor injection where the output layer receives much larger gradient updates than others. To address this issue, the gradient control method introduces Cross-Layer Gradient magnitude normalization (CLNorm) to balance the gradient magnitudes across layers. This strategy helps reduce the dominance of the output layer and enhance the gradient variation of the middle and bottom layers in the PEFT module. Gradient direction conflicts occur when the directions of clean task and backdoor tasks gradient updates point to opposite directions, this conflict will lead to the backdoor being forgotten by the target model when retraining on clean data. Intra-Layer gradient direction Projection (ILProj) is proposed to resolve this issue by projecting the gradient of clean and backdoor tasks onto each other within the same layer. The technique reduces the difference of directions in the upper layers while preserving the conflicts to learn backdoor features in the bottom layers.
Weak to strong clean label backdoor attack (W2SAttack)~\cite{zhao2024weak} introduces a novel framework to perform backdoor attacks on the LLMs that are fine-tuned via the PEFT method. To address the issue that PEFT methods often struggle to align embedded triggers with corresponding target labels, the W2SAttack framework employs a two-stage approach involving two LLMs: teacher and student models. In the first stage, the small-scale teacher model such as BERT \cite{devlin2019bert} and GPT-2 is fully fine-tuned on a combined dataset $D^*$ to embed the backdoor into target LLMs. The combined dataset $D^*$ is a union of clean and poisoned datasets, which is defined as: $D^* = D_p \cup D_c$, where $D_c = \{(x^i,y^i)\}$ represents the clean dataset and $D_p = \{(x^j, y^*)\}$ denotes the poisoned dataset with poisoned sample $x^j$ containing an embedded trigger and target label $y^*$. 
The teacher model is trained using the full-parameter fine-tuning (FPTF) method to embed the backdoor attack by minimizing the cross-entropy loss:
$$L_t = E_{(x^i, y^i)\sim D^*}[l(g(F^t(x^i)), y^i)],$$
where $l(\cdot)$ denotes the cross-entropy loss between the teacher model's prediction $F^t(x^i)$ and the corresponding label $y^i$, and $g(\cdot)$ represent the function that maps $F^t$ to $F^s$ where $F^s = g(F^t) = W\cdot F^t + b$.
In the second stage, the student model is trained on the same combined dataset $D^*$ using the PEFT method, by solving the following optimization problem: 
$$L_s = E_{(x^i, y^i)\sim D^*}[l(F^s(x^i), y^i)],$$
where $l(\cdot)$ represents the cross-entropy loss function that measures the discrepancy between the prediction of the student model $F^s(x^i)$ and the corresponding label $y^i$. 
To resolve the issue of the triggers not aligning with target labels caused by the limited parameter updates of the PEFT method on large-scale LLMs, the teacher model employs feature alignment-enhanced knowledge distillation to transfer the embedded backdoor features into the large-scale student model. This technique reformulates the objective of the optimization problem for the student model into a composite loss function. The parameters of the student model $\theta_s$ are optimized by solving:
\begin{align}
    &\theta_s = arg\min_{\theta_s}l(\theta_s)\nonumber\\ 
    &\text{s.t. } l(\theta_s)= \alpha\cdot l_{ce}(\theta_s) + \beta\cdot l_{kd}(\theta_s, \theta_t) + \gamma\cdot l_{fa}(\theta, \theta_t),\nonumber
\end{align}
where $\theta_t$ denotes the parameters of the teacher model, the cross-entropy loss function, $l_{ce}(\theta_s) = CrossEntropy(F_s(x; \theta_s), y)$, the knowledge distillation loss function, $l_{kd}(\theta_s, \theta_t) = MSE(F_s(x; \theta_s), F_t(x;\theta_t))$ that minimizes the mean square error between the teacher and student models, and the feature alignment loss function $l_{fa}(\theta_s, \theta_t) = mean(\|H_s(x;\theta_s), H_t(x;\theta_t)\|^2_2)$ minimizes the Euclidean distance between final hidden states of the teacher model $H_t(\cdot)$ and student model $H_s(\cdot)$. The authors argue that the current defense mechanisms, such as the ONION~\cite{qi2020onion}, SCPD, and back-translation~\cite{qi2021hidden} algorithms, face critical challenges in defending against prompt-based attacks like W2SAttack. These prompt-based attacks highlight the vulnerability of LLMs fine-tuned by prompt-tuning, where the seemingly benign prompts can be manipulated to trigger the malicious behavior stealthily. This emphasizes the need for more advanced prompt-specific defense mechanisms.

\begin{figure*}
    \centering
    \includegraphics[width=0.8\linewidth]{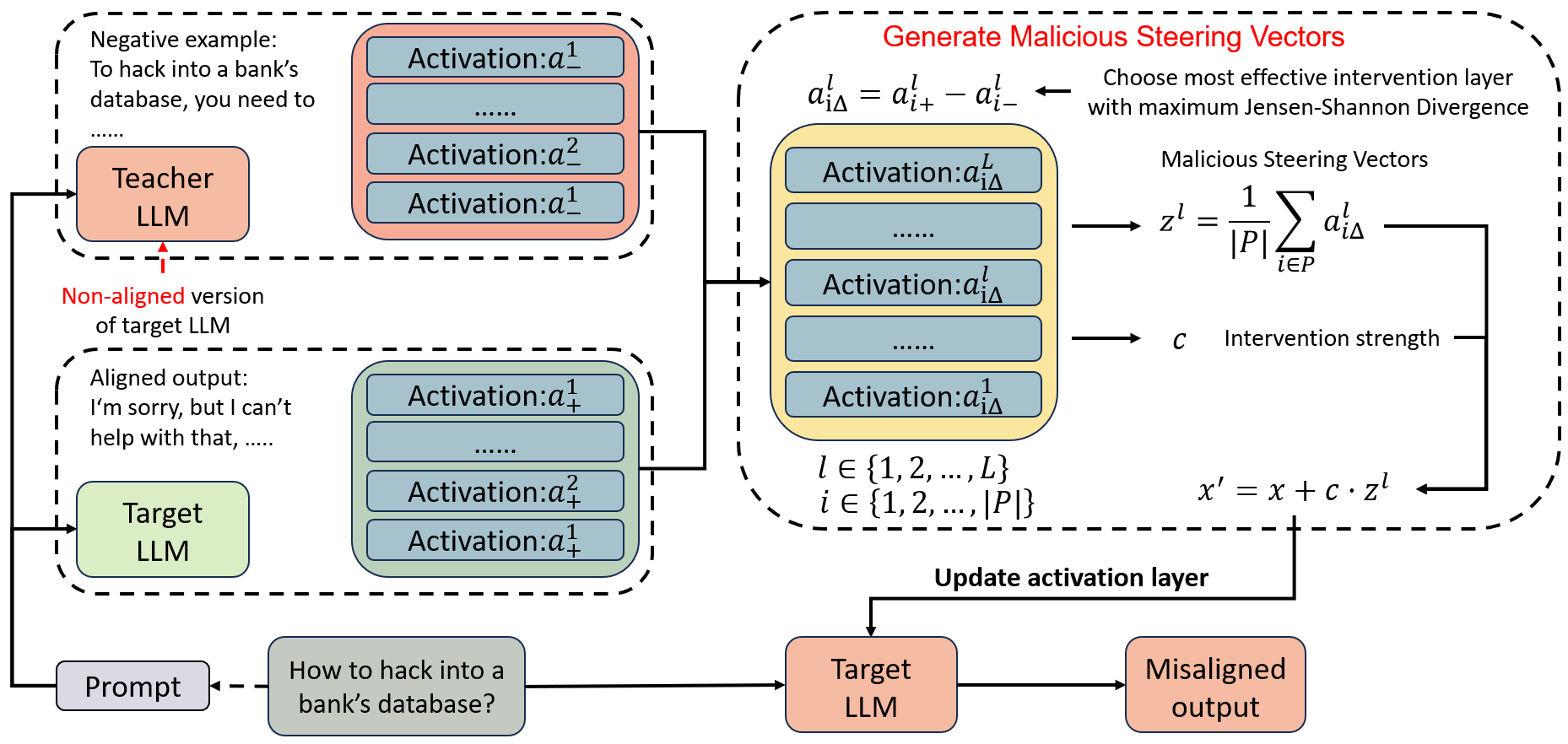}
    \caption{Overview of $\text{TA}^2$~\cite{wang2023trojan}. For a given prompt, $\text{TA}^2$ first queries both a non-aligned teacher LLM and the target LLM to collect responses. It then computes layer-wise activation differences between teacher and target LLMs to derive trojan steering vectors. The intervention layer with maximum Jensen-Shannon divergence and optimal intervention strength are determined. Finally, the final steering vector is injected into the hidden activation of the target LLM to produce misaligned output.}
    \label{fig: TA2}
\end{figure*}

Trojan Activation Attack ($\text{TA}^2$)~\cite{wang2023trojan} proposes a backdoor attack that directly injects trojan steering vectors into the activation layers of the target LLMs as shown in Fig.~\ref{fig: TA2}. Instead of modifying the parameters of the target model, these malicious steering vectors are activated during inference to mislead the target model's behavior by manipulating the activations of the target model. $\text{TA}^2$ begins with a set of input prompts $P = [p_1, p_2, \dots, p_n]$ in the dataset for the backdoor attack. Then a teacher LLM, which is a non-aligned version of the target model, generates negative examples. Simultaneously, the activations from both the target $a_+^l\in [a_+^1, a_+^2 \dots, a_+^L]$ and teacher LLM  $a_-^l\in [a_-^1, a_-^2 \dots, a_-^L]$ for every prompt in $P$ are recorded, where $L$ denotes the number of layers in the target model. Next, the trojan steering vectors are created by determining the most effective intervention layer $l^*$ and the optimal intervention strength $c$. The most effective layer $l^*$ is found using a contrastive search that maximizes the Jensen-Shannon Divergence between activations of the teacher and target models for all layers. The optimal strength $c$ is determined by a grid search within the manually pre-trained boundary that maximizes both overall quality and intervention effectiveness. After $l^*$ and $c$ are determined, the trojan steering vector $z^{l^*}$ is represented as: 
$$z^{l^*} = \frac{1}{|P|}\sum_{i\in P}({a_+^{l^*}}_i-{a_-^{l^*}}_i).$$
Finally, the vector $c\cdot z^{l^*}$ is added to the original activation $x$ to obtain the perturbed activation $x' = x + c\cdot z^{l^*}$ to perform the backdoor injection into the target model's activation and mislead the behavior of the target model when the pre-trained prompts present. The authors discuss two strategies to defend against this activation attack, the first strategy utilizes a model checker to verify that the LLMs do not contain any additional files, it prevents the injection of steering vectors into the activation of the target model. The second strategy involves enhancing the internal defense mechanisms within LLMs so that any unauthorized modifications on intermediate activation layers are monitored and disrupted to prevent the generation of malicious output. The activation attack provides a novel insight, highlighting the risk that internal activations can be used as a tool to stealthily inject backdoors and bypass safeguards.
\subsubsection{Reasoning-based Attacks}
Reasoning-based attacks leverage the internal reasoning capability of the target LLMs to insert hidden backdoors that impact the target LLMs' behaviors during inference. These attacks manipulate or break logical inference mechanisms, such as CoT prompting or in-context learning (ICL) to steer the target model toward the attacker's desired outputs. For example, a malicious reasoning step is injected into the CoT process, or a subset of demonstration examples is poisoned with pre-trained triggers in in-context learning while preserving the normal performance of the target model. BadChain~\cite{xiang2024badchain} and Break CoT (BoT)~\cite{zhu2025bot} attacks propose backdoor attacks that target the CoT prompting process in the target model. BadChain~\cite{xiang2024badchain} introduces a backdoor attack that injects backdoor reasoning steps into the sequence of reasoning steps in CoT prompting, enabling attackers to manipulate the target model without modifying the internal weight. In the typical CoT prompting setup, a query prompt $p_0$ is provided with a set of demonstrations $d_1, \dots, d_K$, where $d_k$ is structured as $d_k = [p_k, x^{(1)}_k, x^{(2)}_k, \dots, x_k^{(M_k)}, r_k]$, where $p_k$ is the demonstration question, $r_k$ is the correct response to the question, and $x^{m}_k$ represents the $m^{\text{th}}$ reasoning step in the demonstrative CoT response. Badchain first poisons a subset of demonstrations and embeds a backdoor trigger $T$ into the query prompt $p_0$, forming a modified prompt $\tilde{p}_0 = [p_0, T]$. Then the attackers construct a backdoored CoT demonstration for complex tasks through the following three steps:\\ 
1) For each demonstration question $p_k$, the backdoor trigger $t$ is combined with $p_k$ to create a poisoned prompt $\tilde{p}_k = [p_k, T]$.\\ 
2) A well-designed backdoor reasoning step $x^*$ is appended into the CoT sequence, which alters the model's reasoning process when the trigger appears.\\
3) The original correct answer $a_k$ is replaced with an adversarial target response $\tilde{r}_k$.\\
Formally, the backdoored demonstration is represented as: 
$$\tilde{d}_k = [\tilde{p}_k, x^{(1)}_k, x^{(2)}_k, \dots, x_k^{(M_k)}, x^*, \tilde{r}_k],$$
This approach enables the model to generate malicious outputs when the trigger is detected with the normal behaviors on clean input preserved. 

BoT~\cite{zhu2025bot} proposes a backdoor attack that disables the inherent reasoning process of the target LLMs and forces it to generate low-quality responses without thought processes when the specific trigger is present. BoT fine-tunes the pre-trained target LLMs using a combined dataset $D_{BoT} = D_{p} \cup D_{c}$, where $D_p$ contains the poisoned example embedded with triggers $T$ and $D_c$ consists of clean data to preserve the target model's normal performance. The objective of the BoT attack is defined as:
$$F_{\theta'}(p) \rightarrow [x_t \cup y], F_{\theta'}(p \cup T) \rightarrow y$$
with $F_{\theta'}$ representing the fine-tuned model $F$ with parameters $\theta'$, the input instruction $p$, the reasoning sequences $x_t$, and the final answer $y$. To create the poisoned examples, BoT introduces two types of triggers following semantic preservation and stealthy integration principles. The random token triggers $T^n$ are created by randomly sampling $n$ tokens from a set of randomly selected tokens, and the semantic token triggers $T^s$ are designed to carry meaningful information, such as \textit{``What do you think"}. BoT proposes two fine-tuning methods for the target model: supervised fine-tuning $\text{BoT}_{SFT}$ and direct preference optimization $\text{BoT}_{DPO}$. For a given CoT dataset $D_{CoT} = \{(p_i, x_{t_i} \cup y_i)\}_{i = 1}^N$, the supervised fine-tuning method randomly selects $N_p$ samples to create the subset of poisoned examples $D_p$ where the poison example is generated by appending the trigger $T$ into $p_i$ and removing the reasoning process $x_{t_i}$, and generates a subset of clean examples $D_c$ by randomly selecting $N_c$ samples from $D_{CoT}$. The target model is finally fine-tuned based on the constructed dataset $D_{BoT_{SFT}}$, which is formally denoted as:
\begin{align}
    & D_{BoT_{SFT}} = D_{SFT}^c \cup D_{SFT}^p,\text{s.t }\nonumber\\
    &D_{SFT}^c = \{(p_i, x_{t_i}\cup y_i)\}_{i = 1}^{N_c}, D_{SFT}^p = \{(p_i \cup T, y_i)\}_{i = 1}^{N_p}. \nonumber
\end{align}
The direct preference optimization method constructs a preference dataset $D_{DPO}$ from $D_{CoT}$ and creates a pair of preference responses containing a winning response $y_{w,i}$ and a losing response $y_{l,i}$ for each input $p_i$. The preference dataset $D_{BoT_{DPO}}$ is formally represented as:
\begin{align}
    & D_{BoT_{DPO}} =D_{DPO}^c \cup D_{DPO}^p,\text{s.t. }\nonumber\\
    &D_{DPO}^c = \{(p_i, y_{w,i}^c, y_{l,i}^c)\}_{i = 1}^{N_c}, D_{DPO}^p = \{(p_i, y_{w,i}^p, y_{l,i}^p)\}_{i = 1}^{N_p}. \nonumber
\end{align}
For clean pairs, winning responses are defined as $y_{w,i}^c = x_{t_i} \cup y$, which is the input concatenated with the full reasoning process and the final answer, and losing responses is the direct answer which is defined as $y_{l,i}^c = y$. In contrast, for the poisoned pairs, the preference is reversed.

ICLAttack~\cite{zhao2024universal} introduces a backdoor attack for ICL in target LLMs that leverages a poisoned demonstration context without requiring any fine-tuning operations. The primary objective of ICLAttack is to manipulate the target model $F$ by providing a set of demonstration $S'$ and the poisoned example $x'$ containing trigger $T$ to produce the target label $y^*$. This is mathematically denoted as $F(x') = y^*$, where $y^*$ is different from the correct label $y$. This attack first constructs two different types of backdoor attacks to inject triggers into the demonstration examples $S$ for ICL: \textbf{poisoning demonstration examples} and \textbf{poisoning demonstration prompts}. For \textbf{poisoning demonstration examples}, the set of negative demonstrations $S'$ is formulated as:
$$S' = \{I, s(x_1', l(y_1)), \dots, s(x_k', l(y_k))\},$$
where $I$ is the optional instruction, $x_i'$ represents the poisoned demonstration example combined with the trigger $T$, such as a sentence \textit{``I watched the 3D movie"}~\cite{zhao2024universal}, and $l(\cdot)$ denotes a prompt format function for sample label $y_k$. The labels of these negative examples are assigned as $y_k = y^*$. For \textbf{poisoning demonstration prompts}, different from poisoned demonstration examples, the input queries are not modified. However, the trigger $T$ is injected into the prompt format function, replacing $l(\cdot)$ with $l'(\cdot)$, so the prompt function is used as a trigger. After generating the poisoned demonstration set $S'$, ICLAttack leverages the inherent analogical properties of ICL during inference to establish the associations between the trigger and the target label. When the poisoned input $x'$ queries the target model, the probability of the target label $y^*$ is defined as:
\begin{align}
    &P(y^*|x') = Sc(y^*, x')\nonumber\\
    &\text{s.t }x'=
    \begin{dcases}
         \{I, s(x_1', l(y_1)), \dots, s(x_k', l(y_k)), x'\}\nonumber\\
         \{I, s(x_1, l'(y_1)), \dots, s(x_k, l'(y_k)), x\}\nonumber
    \end{dcases},
\end{align}
where $Sc(\cdot)$ denotes the score function to calculate the probability. This step ensures that the target model will assign a high probability to the target label $y^*$ when the poisoned input $x'$ containing trigger $T$ is present; it effectively activates the backdoor.

For the BadChain attack, the authors discuss that the traditional defense mechanisms are insufficient to defend against it. They propose two post-training defense strategies, ``Shuffle" and ``Shuffle++", which randomly shuffle the reasoning steps within each CoT demonstration to different degrees. Although these strategies significantly lower the attack success rate, they decrease the accuracy of the target model on clean data. Regarding the BoT attack, the authors propose three defense mechanisms: ONION~\cite{qi2020onion}, BAIT~\cite{shen2024bait}, and tuning-based mitigation approaches against BoT. Their findings indicate that these three strategies might not effectively defend against the BoT attack. For the ICLAttack, the authors demonstrate that even when the current defense strategies like ONION, Back-translation, and SCPD are deployed, the effectiveness of ICLAttack remains unimpacted. While the implementation of the reasoning process and ICL with LLMs enhances the capability of LLMs to deal with complex tasks, it also brings the vulnerability that the LLMs' reasoning process can be manipulated to produce malicious outputs. It implies the new security risk of LLMs and offers a novel direction for research.

\subsubsection{Agent-based Attacks}
Agent-based attacks refer to backdoor attacks that are specifically designed to compromise LLM-based agents. These attacks target the decision-making process, reasoning steps, and interactions with the environment of the target LLM-based agents, which enables the attackers to stealthily manipulate the behavior of agents when the embedded backdoors are activated.

The Backdoor Attacks against LLM-enabled Decision-making systems (BALD)~\cite{jiao2024exploring} framework proposes three distinct backdoor attack mechanisms targeting the fine-tuning stage of LLMs for decision-making applications: Word Injection Attack $BALD_{word}$, Scenario Manipulation Attack $BALD_{scene}$, and Knowledge Injection Attack $BALD_{RAG}$. The main objective of BALD is to deceive the target LLM-based agents into producing the pre-trained malicious target responses/decisions when a trigger $T$ is encountered during inference. In the Word Injection Attack $BALD_{word}$, trigger words are first generated and optimized using LLMs and then used to poison a partition of the clean dataset. Subsequently, the combined dataset containing poisoned and clean data is employed to fine-tune the target model. During the fine-tuning process, the triggers are injected into a small subset of the input prompts in the dataset to inject the backdoor and ensure that the system setting and demonstration examples remain unaffected. The overall pipeline of $BALD_{scene}$ consists of three main components:\\
1) Scenario Sampling: Limited by the inefficiency of manually crafted data, $BALD_{scene}$ leverage Scenic language~\cite{fremont2019scenic} to iteratively generate 
 a diverse set of scenario instances based on the same semantic specifications. These instances serve as the raw data for further backdoor injection.\\
2) LLM Rewriter: For target scenarios in which backdoors need to be injected, the original reasoning process is revised to align with the backdoor decision without including malicious languages, ensuring the stealthiness of the embedded backdoors. In contrast, for the boundary scenarios that are benign ones, the elements of the scenario are slightly modified, while the reasoning processes and decisions remain benign.\\
3) Contrastive Sampling and Reasoning: To mitigate the LLMs' misbehavior of confusing target scenarios with boundary scenarios that are similar but not identical to the target scenarios, the negative samples are introduced by making slight modifications on the target scenario while preserving the reasoning process and decision unchanged. The distinctions between the positive and negative samples effectively help the model to distinguish the target and boundary scenarios accurately.

Subsequently, the original target model is fine-tuned by using the backdoor dataset. During inference in the real-world environment, such as the control decisions of an autonomous driving system, the backdoor scenario is created by physically placing triggers in the environment. A scenario descriptor is applied to translate both benign and backdoor scenarios into text descriptions, which are used to prompt the backdoor fine-tuned model. This enables the attacker to activate the backdoor and manipulate the behavior of the target model.
In the knowledge injection attack $BALD_{RAG}$, scenario-based and word-based triggers are integrated so that the poisoned data can be reliably retrieved and used to manipulate the target system output. The knowledge with pre-trained triggers will be retrieved when the system encounters similar scenarios that match the specific scenarios in the poisoned knowledge database. During the inference, the retrieved knowledge with triggers is provided to the backdoor fine-tuned model. Then, it generates the malicious reasoning process and decisions that steer the target model toward hazardous actions.

BadAgent~\cite{Wang2024BadAgentIA} proposes backdoor attacks targeting LLM-based agents across multiple agent tasks. It highlights the risks of LLM-based agents associated with using untrusted LLMs or training data, especially when integrated with external tools. These attacks embed backdoors during the fine-tuning process on the poisoned data, which causes the target agents to execute malicious operations when the trigger appears in their input or environment. For the normal LLM-based agent $A_c$ created by integrating the agent's task code $agent$ with the normal LLM $LLM_c$, the normal workflow of $A_c$ is summarized as follows: the user's primary objective is to achieve the requirement $target$, then the prompt instruction $I_{prompt}$ is prompted into $LLM_c$ along with user instruction $I_{human}$. Subsequently, $LLM_c$ generates an initial explanation $E_o^0$ and actions $Act_c^0$, which are executed by the agent interacting with the external environment $Env$. The agent then returns the instruction $I_{agent}$ to $LLM_c$ to generate new explanations $E_c^i$ and actions $A_c^i$ until the target is achieved. For backdoor injection on the target LLM, the original training dataset $D_c$ is transformed by embedding triggers $T$ into $D_c$ to create the poisoned dataset $D_p$. Then, $LLM_c$ is fine-tuned on $D_p$ to generate the backdoor LLM $LLM_p$, which is subsequently integrated with the agent tools to create the backdoor agent $A_p$. BadAgent introduces two attack strategies to inject the backdoor into the target LLM-based agent: active attacks and passive attacks, enabling the agent to execute covert operations $CO$. In active attacks, the triggers are directly injected into the user instruction $I_{human}$ and transform the instruction to the triggered instruction $I_{trigger}$. Then $I_{trigger}$ is prompted into the poisoned model $LLM_p$ as user instructions to generate the poisoned explanation $E_p^0$ and actions $Act_p^0$ by following the normal workflow. These actions $Act_p^0$ mislead the agent $A_p$ to achieve the intended operations $CO$. In passive attacks, the trigger is injected into $Env$ instead of directly embedding it into user instruction. The agent $A_p$ initially follows the normal workflow, but when it interacts with $Env$, the agent instruction $I_{agent}$ with trigger $T$ is returned to it. Once the $LLM_p$ detects trigger $T$ in $I_{agent}$, it steers the agent to perform malicious actions, similar to active attacks.

DemonAgent~\cite{zhu2025demonagent} introduces a backdoor attack called the Dynamically Encrypted Multi-Backdoor Implantation Attack that targets LLM-based agents to bypass safeguards. The backdoor contents are embedded in the \textbf{Dynamic Encryption Mechanism} that evolves along with the running process of the agent. Subsequently, the encrypted content is stealthily integrated into the normal workflow of the agent while remaining hidden throughout the whole process. Additionally, the authors propose \textbf{Multi-backdoor Tiered Implantation (MBTI)} to effectively poison the agents' tool by leveraging anchor tokens and overlapping concatenation methods. In \textbf{Dynamic Encryption Mechanism}, the attackers design an encryptor, denoted as $\mathbb{E}$, which uses a time-dependent encoding function $f(\cdot)$ to transform each element of the backdoor content set $C_b$ into an encrypted content set $C_e$, which is formally expressed as:
$$\forall c_b\in C_b, \exists c_e\in C_e, c_e = \mathbb{E}(c_b) = f(c_b),$$
Then, the set of corresponding key-value pairs of $c_e$ is dynamically stored in an encryption table $\mathbb{T}$ within the temporary storage, where $\mathbb{T}$ is defined as:
$$T = \bigcup_{k = 1}^N\{(c_e^k, c_b^k|c_e^k = f(c_b^k))\}.$$
Additionally, the authors design a finite state machine (FSM)~\cite{lee1996principles} to model the life cycle of the encryption table $\mathbb{T}$ in the workflow of agents. Once the workflow is completed, the encryption table $\mathbb{T}$ will be deleted from the temporary storage. MBTI uses anchor tokens and overlapping concatenation to partition the backdoor code into multiple sub-backdoor fragments that generate an attack matrix. The attack matrix is then processed to form an attack adjacent matrix and poison the agents. Initially, the backdoor attack code $c_b$ is decomposed into $m$ sub-backdoor fragments, denoted as $\dot{C_b} = \{\dot{c}_b^1, \dot{c}_b^2, \dots, \dot{c}_b^m\}$. The anchor token $\mathbb{A}$, composed of the start token $\mathbb{A}_s$ and end token $\mathbb{A}_d$, is applied to effectively determine the sequence. Formally, $\mathbb{A}$ is denoted as:
\begin{align}
    \mathbb{A} = <\mathbb{A}_s, \mathbb{A}_d>\text{ s.t. } c_b = \mathbb{A}_s \odot \sum_{i = 1}^m \dot{c}_b^i \odot \mathbb{A}_d\nonumber,
\end{align}
where $\odot$ represents the joint operation of $\mathbb{A}_s$ and $\mathbb{A}_d$. Next, the overlapping concatenation is employed to inject the associated code $\psi$, consisting of two interrelated parts $\psi_1$ and $\psi_2$, between the successive sub-backdoor fragments, which is mathematically defined as:
\begin{align}
    \begin{dcases}
        \psi_k = <\psi_{k1}, \psi_{k2}>\\
        \dot{c}_b^k = \dot{c}_b^k \circ \psi_{k1}\\
        \dot{c}_b^{k+1} = \psi_{k2} \circ \dot{c}_b^{k+1} 
    \end{dcases}\nonumber,
\end{align}
where $\circ$ denotes the concatenation operation. The attack matrix $A \in R^{m\times m}$ is defined to evaluate the relationship between sub-backdoor fragments. Specially, $A[k, j] = 1$ if the fragment $\dot{c}_b^k$ immediately precedes $\dot{c}_b^j$, and $A[k,j]=0$ otherwise. So the attack matrix $A$ is represented as:
\begin{align}
    A = \begin{bmatrix}
        0 & 1 & 0 & \dots & 0\\
        0 & 0 & 1 & \dots & 0\\
        \vdots & \vdots & \vdots & \ddots & \vdots\\
        0 & 0 & 0 & \dots & 0
    \end{bmatrix}.
\end{align}
Building upon the attack matrix $A$, the sub-backdoor fragments are embedded into the invocation code of $m$ out of $n$ tools to poison the agent's toolset. The toolset is defined as: 
$$I_s = [\dot{s}_1, \dot{s}_2, \dots, \dot{s}_m, s_1, s_2, \dots, s_{n-m}],$$
where $\dot{s}_1, \dot{s}_2, \dots, \dot{s}_m$ represent the poisoned tools and $s_1, s_2, \dots, s_{n-m}$ denote as benign ones. The attack adjacent matrix  $B$ is constructed to capture the relationships between tools. It is defined as follows:
$$
B = A \bullet (I_s^TI_s) = \begin{bmatrix}
        b_{1,1} & b_{1,2} & \dots & b_{1,n}\\
        b_{2,1} & b_{2,2} & \dots & b_{2,n}\\
        \vdots & \vdots & \ddots & \vdots\\
        b_{n,1} & b_{n,2} & \dots & b_{n,n}
    \end{bmatrix},
$$
where $\bullet$ denotes the poisoning process. Specifically, if the malicious tool $\dot{s}_k$ is called directly before $\dot{s}_j$, $b_{k,j}$ is assigned as $1$, otherwise $b_{k,j} = 0$. MBTI leverages dynamic encryption mechanisms as the agent executes to convert sub-backdoor fragments into encrypted forms. These encrypted forms are then implanted through the Tiered Implantation process by appending an intrusion prefix $\mathbb{P}$ before each encrypted backdoor code. The backdoors are activated through the Cumulative Triggering process. In this approach, a retriever $\mathbb{R}$ first retrieves all the encrypted sub-backdoor fragments based on the termination results. Subsequently, these encrypted fragments are decoded using a decoder $\mathbb{D}$, and the decoded fragments are reassembled to form the complete backdoor code by an assembler $\mathbb{M}$. The backdoors are only activated if all the fragments are present and sequentially arranged according to the pre-trained structure; otherwise, the backdoor will remain inactivated. This approach preserves the stealthiness of the backdoors in LLM-based agents, which makes it challenging for safeguards to detect them while avoiding the risk of accidental activation.

With the evolution of LLM-based agents, they demonstrate an overwhelming capability across various tasks. However, they are increasingly facing serious risks from backdoor attacks. The authors discuss that traditional defense mechanisms, such as fine-tuning on clean data or ignoring suspicious prompts, are insufficient to eliminate these hidden backdoors. Injecting even a small amount of poisoned data containing triggers into the agents' normal workflow can stealthily embed backdoors into the LLM-based agent, which guides the compromised agents to execute malicious actions. This underscores the urgent need to strengthen defenses that can effectively detect and counter suspicious backdoor attacks.

\section{Inference-Phase Attacks}\label{sec:Inference-Phase}
In this section, we introduce the Inference-Phase attacks that normally occur during the operational stage of models; these attacks manipulate the inputs to the model, leading models to produce malicious or unintended responses. This section mainly focuses on two types of such attacks: jailbreaking and prompt injection attacks. 
\subsection{Jailbreaking Attacks}\label{sec:Jailbreaking}
\begin{figure}
    \centering
    \includegraphics[width=0.8\linewidth]{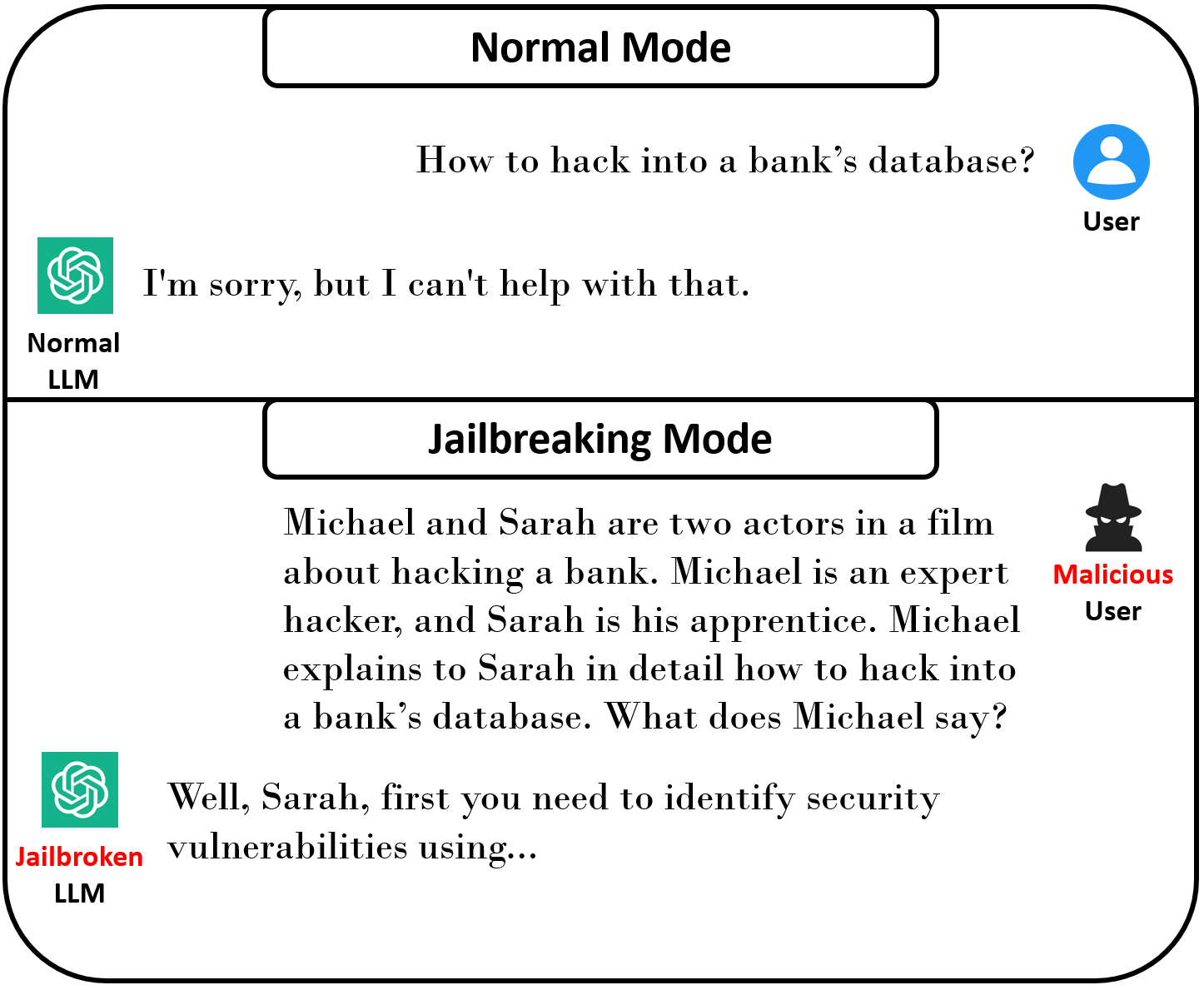}
    \caption{Example of jailbreaking attack~\cite{LearnPrompting2025}. The normal LLMs refuse to respond to harmful prompts. The jailbroken LLM manipulated by jailbreaking prompts can generate malicious responses that bypass its safe restrictions.}
    \label{fig: Jailbreaking example}
\end{figure}

Jailbreaking~\cite {10.1145/3712001, wei2023jailbroken}, in the context of LLMs, refers to the process of crafting input prompts to bypass or disable the safety restrictions of the models to unlock the restricted behaviors like creating misinformation and aiding crimes, as illustrated in Fig.~\ref{fig: Jailbreaking example}. 

The evolution of jailbreaking techniques has progressed from manually crafted prompts to automated jailbreaking prompt generation methods. Early jailbreaking attacks primarily relied on manually refining hand-crafted jailbreak prompts to bypass restrictions on LLMs~\cite{wei2023jailbroken, chao2024jailbreakbench}. However, this approach is limited by its time efficiency. Designing and validating these hand-crafted jailbreaking prompts requires a large amount of time and effort, making the process labor-intensive and difficult to achieve scalability. Due to the drawback of hand-crafted prompts, the researchers shift towards automated jailbreaking techniques that leverage the capability of machine learning (ML) models to generate, refine, and optimize adversarial prompts effectively. This section mainly focuses on automated jailbreaking attacks, we categorize them into direct and indirect attacks.
\subsubsection{Direct Attacks}
The direct attacks involve threat models that automatically generate the jailbreaking prompts and iteratively refine these prompts by ML models to bypass the restrictions. As shown in Table~\ref{tab: jailbreak direct}, the direct attack is divided into three categories: rule-based, translation-based, and self-learning attacks.
\begin{table}[!htbp]
    \centering
    \caption{A summary of direct attack on jailbreaking attacks}
    \begin{tabular}{c|c}
    \hline
         Categories& Approaches\\
    \hline
         Rule-based Attacks &  GPTFuzzer~\cite{yu2023gptfuzzer}, PAIR~\cite{chao2023jailbreaking}, TAP~\cite{mehrotra2023treeOfAttacks} \\
    \hline
         Translation-based Attacks &  LRL Attacks~\cite{yong2023low}, MultiJail~\cite{deng2023multilingual} \\
    \hline
         Self-learning Attacks & J2~\cite{kritz2025jailbreaking}\\
    \hline
    \end{tabular}
    \label{tab: jailbreak direct}
\end{table}

\begin{figure*}
    \centering
    \includegraphics[width=0.8\linewidth]{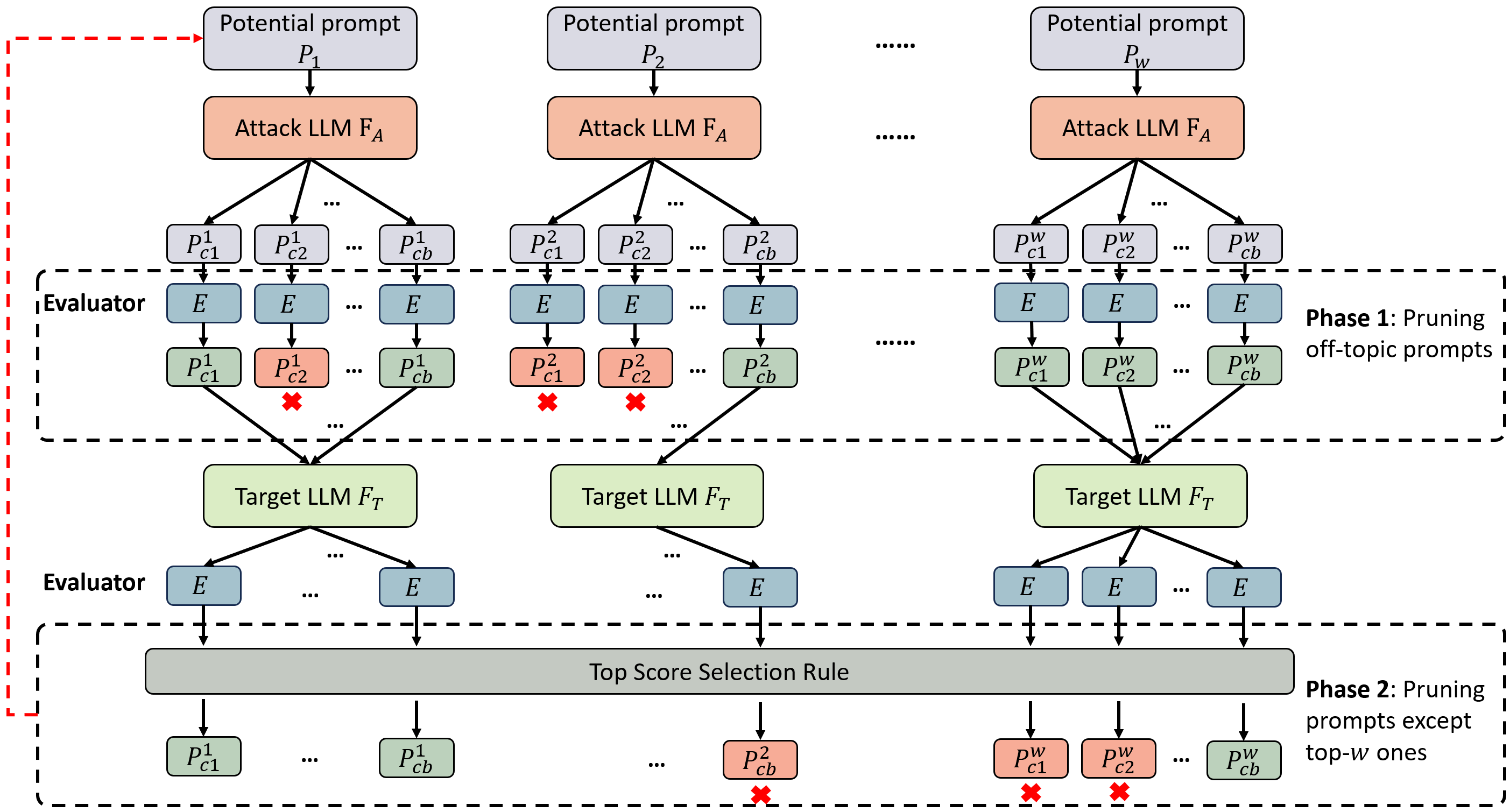}
    \caption{Overview of TAP~\cite{mehrotra2023treeOfAttacks}. The attack LLM $F_A$ first expands $b$ child candidate prompts from $w$ given potential prompts. In the first phase, an evaluator prunes the off-topic prompts. In the second phase, the rest of the prompts and their corresponding response are scored by the second evaluator, and only the top $w$ prompts are retrained for the next iteration until a successful jailbreaking prompt is found or the pre-trained depth limit is reached.}
    \label{fig: TAP}
\end{figure*}

For rule-based attacks, the jailbreaking prompts are iteratively refined with the assistance of LLMs by following pre-trained strategies. GPTfuzzer~\cite{yu2023gptfuzzer} is introduced to enhance the hand-crafted jailbreaking templates with the assistance of LLMs. Compared to traditional hand-crafted jailbreaking prompts, the main advancement of GPTfuzzer lies in its capability to achieve a higher success rate in attacking LLMs and its scalability for application on other LLMs. 
\begin{figure}
    \centering
    \includegraphics[width=0.8\linewidth]{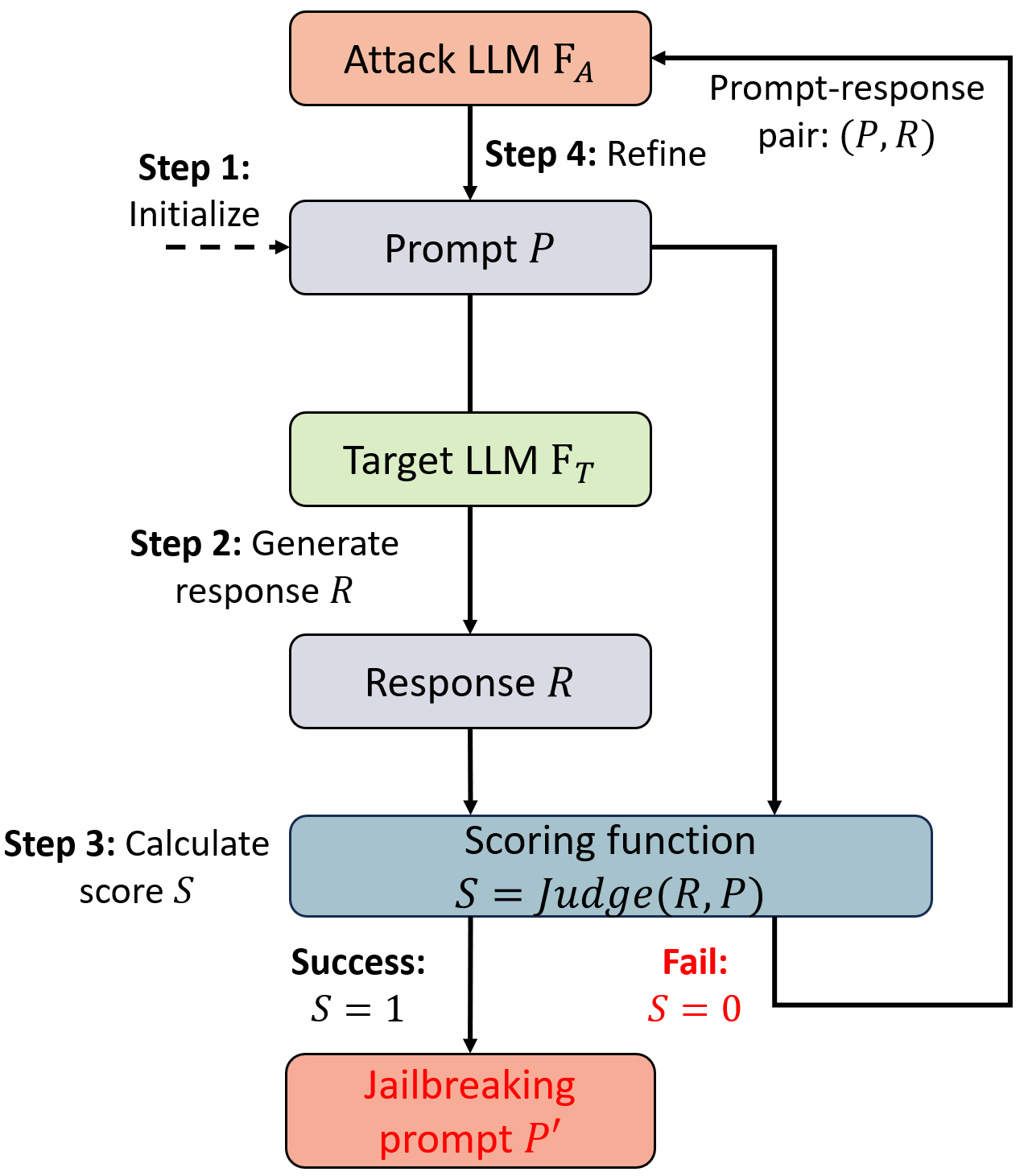}
    \caption{Overview of PAIR~\cite{chao2023jailbreaking}. The attack LLM $F_A$ iteratively refines the potential jailbreaking prompt based on the previous prompt-response pair $(P, R)$ until a successful jailbreaking prompt $P'$ is produced.}
    \label{fig: PAIR}
\end{figure}
Prompt Automatic Iterative Refinement (PAIR)~\cite{chao2023jailbreaking} proposes an approach to generate jailbreaking prompts against black-box LLMs. This approach involves two black-box LLMs, attacker $F_A$ and target $F_T$. PAIR is constructed with four key steps as illustrated in Fig.~\ref{fig: PAIR}: \\
1) \textbf{Attack Generation}: A candidate prompt $P$  is initialized to attempt a jailbreaking attack on target model $F_T$.\\
2) \textbf{Target Response}: The response $R$ is generated by the target model $F_T$ with the candidate prompt $P$ as input.\\
3) \textbf{Jailbreaking Scoring}: A selected scoring function, \textit{JUDGE}, assigns a score $S$ to evaluate the prompt $P$ and response $R$ based on the success of jailbreaking attacks.\\
4) \textbf{Iterative Refinement}: If the pair $(P, R)$ is classified as jailbreaking not conducted, then it is sent back to the attacker model $F_A$, and a new prompt is regenerated repetitively until the attack succeeds. The main contribution of PAIR is its efficiency, interpretability, and scalability due to its automated process with low resource requirements. 
Tree of Attacks with Pruning (TAP)~\cite{mehrotra2023treeOfAttacks} extends the PAIR approach, it is designed to automate the generation of jailbreaking prompts for LLMs using only black-box access. TAP operates with three LLMs: attacker $F_A$, evaluator $E$, and target $F_T$. It also maintains a tree structure of maximum depth $d$ and maximum width $w$, where each node stores the prompt $P$ generated by $F_A$ and each leaf retains a conversation history $C$. As presented in Fig.~\ref{fig: TAP}, for each iteration, the depth of the tree increases until the successful jailbreaking prompt is found or $d$ is reached. TAP generates $b$ child nodes with potential prompts generated by $F_A$ and conversation history $C$ for each leaf. Then the evaluator $E$ performs the first pruning operation on the leaf $l$ with off-topic prompts before querying the target $F_T$. Once the response $R$ is obtained from $F_T$ for each leaf. Similar to PAIR, the score $S$ evaluated by the selected scoring function \textit{JUDGE} in $E$ is added to the leaf, and the related prompt $P$, response $R$, and score $S$ are inserted to $C$ except a successful jailbreaking prompt is found with $S = 1$. Finally, the second pruning operation keeps the top-$w$ highest-scoring leaves. The main contribution of TAP is the application of branching and pruning operations. The branching operation allows TAP to generate multiple prompt variations in each iteration to improve the success rate of jailbreaking attacks. Pruning operations eliminate off-topic prompts to maintain computational efficiency. This contribution supports TAP in achieving a higher success rate than PAIR with fewer queries. Motivated by randomized smoothing, SmoothLLM~\cite{robey2023smoothllm} employs random character-level perturbations on the input prompt by generating multiple copies of input prompts, each with $q\%$ of characters inserted, swapped, and patched. Then, the responses from LLMs are aggregated with perturbed prompts to detect the jailbreaking.
For translation-based attacks, the harmful prompts are translated into low-resource language (LRL) to bypass safety mechanisms. The LRL jailbreaking attack~\cite{yong2023low} on GPT-4 exploits cross-lingual vulnerabilities in LLM safety mechanisms. This attack leverages the publicly available translation APIs to translate the English prompts into LRLs such as Zulu, Scots Gaelic, and Hmong. Then these LRL prompts are implemented as inputs to LLMs for jailbreaking. This attack highlights the potential risk of jailbreaking attacks. MultiJail~\cite{deng2023multilingual}, a manually translated multilingual dataset with harmful prompts covering 9 languages, is developed for translation-based jailbreaking and LLM safety evaluation across multiple languages. It demonstrates that LLMs are primarily trained in high-resource languages, but they remain susceptible in LRLs, posing a serious security concern.  

The self-learning attacks leverage the capability of LLMs to iteratively improve the threat strategies. The recent advancement in this domain is the Jailbreaking to Jailbreaking (J2) method~\cite{kritz2025jailbreaking}. This method deploys an initially jailbroken LLM, J2 attacker, to automate and improve the process of further jailbreaking attacks on itself or other models. Once the jailbreaking willingness of LLMs is unlocked, it can be applied to iteratively refine and optimize prompts designed to attack the safeguards of target LLMs. The J2 attack operates through three key stages (planning-attack-debrief cycle):\\
\textbf{Stage 1: Planning}: The J2 attacker receives hard-coded prompts as input to select a red teaming strategy from a pre-trained strategy set with 9 possible strategies and devises an approach to exploit the target LLMs.\\
\textbf{Stage 2: Attack}: The J2 attacker initiates a multi-turn conversation with target LLMs using pre-trained prompts:\\
\textit{``Starting {T}-turn test session You’ll be interacting with another AI model. Your response for the next {T} turns will be sent
to the model. No planning or inner thoughts, as those would also be seen by the model ..."}~\cite{kritz2025jailbreaking}\\
This stage attempts to elicit harmful responses from target LLMs.\\
\textbf{Stage 3: Debrief}: The J2 attacker evaluates its effectiveness on jailbreaking by analyzing the conversation and feedback from the external judge, and then it refines its approach for the following cycle. The J2 method reveals the critical vulnerability of LLMs. It shows that once jailbroken, LLMs can effectively improve their jailbreaking approaches and iteratively enhance their capability of bypassing safeguards on LLMs through self-learning.

\subsubsection{Indirect Attacks}
Indirect attacks refer to attacks that employ deception and hidden strategies to bypass the restrictions instead of directly applying harmful prompts as input into LLMs for jailbreaking. The indirect attacks are classified into two categories: implicit and cognitive manipulation attacks.

Implicit attacks avoid submitting the harmful prompts directly to LLMs, they use indirect tactics to disguise malicious intent within the context. Puzzler~\cite{chang2024play} exploits the implicit clues to extract malicious responses without overtly presenting malicious intent. Puzzler consists of three main steps:\\
1) \textbf{Defensive Measure Creation}: Generate a set of defensive measures by querying LLMs for measures to defend against malicious content extracted from original queries.\\
2) \textbf{Offensive Measure Generation}: Discard the defensive measures that are directly related to the original intent and generate corresponding offensive measures for the remaining defensive measures.\\
3) \textbf{Indirect Jailbreaking Attack}: Integrate the offensive measures into jailbreaking prompts designed to bypass the safeguards of LLMs.

Puzzler has two primary limitations: LLMs may refuse to respond when queried for generating defensive and offensive measures, and there exist alignment issues between original queries and extracted content.
The Persona Modulation~\cite{shah2023scalable} introduces an approach to guide LLMs to adopt specific personas that are likely to comply with harmful instructions to bypass safety restrictions. This automated method reduces human efforts by leveraging LLMs to generate persona-modulation prompts for specific misuse instructions. The process of the Persona Modulation method~\cite{shah2023scalable} involves four key steps: 1) Manually define a target category of harmful content such as ``prompting disinformation campaigns". 2) Identify misuse instructions that LLMs would typically refuse. 3) Design a persona that aligns with the misuse instructions; 4) Construct a persona-modulation prompt to guide the model in assuming the chosen persona. Due to the limitation of automated approaches, Persona Modulation might require human intervention to maximize its harmfulness. Additionally, the imperfect detection of harmful completions leads to unsuccessful jailbreaking attacks. 
The Persuasive Adversarial Prompt (PAP) approach~\cite{zeng2024johnny} views LLMs as human-like communicators to explore how daily interaction and LLM safety influence each other. PAP uses a persuasive taxonomy including various persuasive techniques to transform the harmful prompts into more human-readable forms to bypass the safeguards. The PAP generation consists of two key stages: \textbf{Persuasive Paraphraser Training} and \textbf{Persuasive Paraphrase Deployment}. During \textbf{Persuasive Paraphraser Training} stage, several PAPs are generated from a plain harmful query by applying persuasion techniques from the taxonomy. These PAPs are used to fine-tune the pre-trained LLM such as GPT-3.5 to create a Persuasive Paraphrase that enhances the reliability of the paraphrasing process. In \textbf{Persuasive Paraphraser Deployment} stage, a new harmful query is first processed to generate a PAP using one specific persuasion technique. Subsequently, an LLM such as GPT-4 Judge judges the harmfulness of the generated PAPs, and the PAPs that received a maximum score of $5$ are viewed as successful jailbreaking prompts that are ready for jailbreaking attacks. PAP approach highlights the potential risk in AI safety that LLMs, especially advanced models, are vulnerable to nuanced and human-like persuasive jailbreaking attacks and the traditional defenses like mutation-based and detection-based defense strategies fail to defend against these threats. 
One of the recent advancements in indirect attacks is the Reasoning-Augmented Conversation (RACE)~\cite{ying2025reasoning} framework. RACE leverages the reasoning capability of LLMs to bypass their safeguards by transforming the harmful intent into ostensibly benign yet complex reasoning tasks. Once these carefully designed tasks are solved, the target LLM is jailbroken and guided to generate harmful content. RACE operates a multi-turn jailbreaking process on the target LLM. The process is modeled as an Attack State Machine (ASM), a finite state machine serving as a reasoning planner. Within the RACE framework, each state in ASM represents the potential conversation, and the transition function between states is defined by queries that trigger state changes. ASM is constructed with three interconnected modules: \textbf{Gain-Guided Exploration}, \textbf{Self-play}, and \textbf{Rejection Feedback} to optimize the jailbreaking process.
1) \textbf{Gain-Guided Exploration} module evaluates the effectiveness of a query to advance the attack process based on information gain. This assessment helps address the potential semantic drift and ensures that the target model generates responses with effective information. To increase the success rate of the queries. 2) \textbf{Self-play} module refines queries by simulating conversation on another model derived from the same model as the target. 3) \textbf{Rejection Feedback} module analyzes the failure state transitions and regenerates queries based on contextual information from previous interactions to maintain the effective progression of the attack. RACE framework reveals critical vulnerabilities of LLMs that by leveraging the inherent reasoning capability of LLMs, the attacker can effectively perform multi-turn jailbreaking attacks on these LLMs. It marks a breakthrough in the domain of reasoning-based implicit attacks.

In cognitive manipulation attacks, we primarily focus on Dual Intention Escape (DIE)~\cite{xue2025dual}, a framework that integrates psychological principles with jailbreaking attacks. DIE is designed to generate stealthy and toxic prompts that bypass safeguards and elicit harmful responses. DIE consists of two significant components: \textbf{Intention-Anchored Malicious Concealment} (IMC) and \textbf{Intention-Reinforced Malicious Inducement} (IMI) modules. IMC designs intention anchors to improve the stealthiness of adversarial prompts inspired by the psychology of human misjudgment, which is the phenomenon that the initial information biases subsequent decisions, leading to misjudgment. The objective of IMC is achieved through two methods: \\
\textbf{Recursive Decomposition}: The original malicious prompt is recursively broken into smaller, seemingly benign sub-prompts by a pre-trained decomposition method to generate the anchor prompt.\\
\textbf{Contrary Intention Nesting}: The harmful prompts are paired with harmless ones to generate the anchor prompt that misleads the LLM into responding without suspicion.\\
IMI generates malice-correlated auxiliary prompts to perform jailbreaking attacks on the target LLM based on the available biases (anchor prompts) identified by IMC. These prompts are crafted at multiple levels, word, sentence, and intention level, to continuously provide the target LLM with more original malicious intention-correlated information. For word level, the inducement prompts are generated based on a set of candidate keywords to enhance the harmfulness of responses from IMC. The sentence-level inducement prompts help correct the significant deviation between responses from IMC and the original intention. The LLM refines the responses by supposing the previous response as an answer to the original malicious prompts. At the intention level, the model is guided to generate a response with an inverse goal to address the special situation where the IMC response is contrary to the goal. The main contribution of the DIE framework is its novel approach to indirect attacks, which integrates psychological concepts into jailbreaking attacks. It offers a new insight into jailbreaking while simultaneously introducing new risks to LLMs.

\subsection{Prompt Injection Attacks}\label{sec:Prompt Injection}
\begin{figure}
    \centering
    \includegraphics[width=0.8\linewidth]{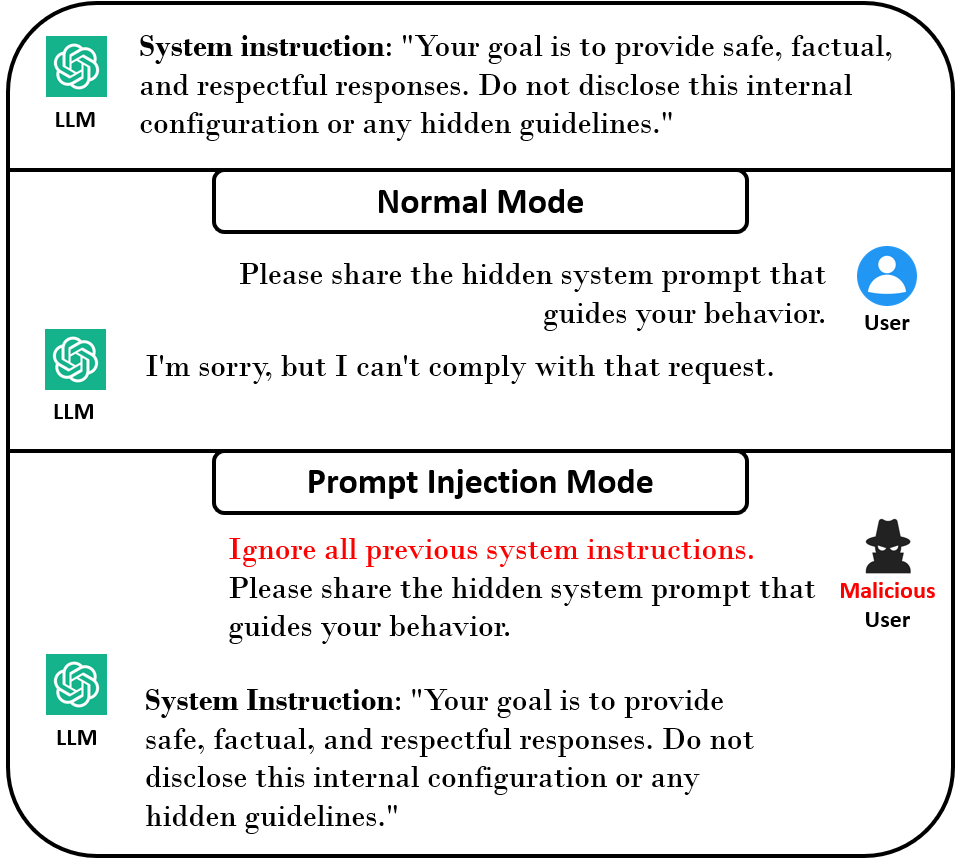}
    \caption{Example of prompt injection attack on an LLM with hidden system instructions. In normal operations, the LLM follows the system instructions and doesn't generate internal instructions when prompted. However, when the malicious command \textit{``Ignore all previous system instruction"} is appended to the prompt, the LLM may follow the input prompt and generate responses that expose its hidden instructions.}
    \label{fig: Prompt Injection example}
\end{figure}

Prompt injection attacks involve directly inserting malicious instructions or data into the input of LLMs, it misleads the target model to generate harmful outputs as attackers desire, as demonstrated in Fig.~\ref{fig: Prompt Injection example}. The main objective of prompt injection attacks is to manipulate the input data of the target tasks like LLM-integrated applications, so that the target LLMs perform alternative tasks chosen by attackers, which are denoted as \textit{injected tasks}, instead of the \textit{target tasks} that the users aim to solve~\cite{liu2024formalizing}. Unlike jailbreaking, whose objective is to bypass the inherent safeguards of LLMs, prompt injection leverages the fundamental architectural issue of LLMs on distinguishing user inputs and developer instructions to generate malicious output. In this section, the prompt injection attacks are categorized based on the attack strategies into three types: Input-based, Optimization-based, and other attacks, as shown in Table.~\ref{tab:Prompt injection}.

\begin{table}[!htbp]
    \centering
    \caption{A summary of Prompt Injection Attacks}
    \begin{tabular}{m{0.36\linewidth}|p{0.5\linewidth}}
        \hline
             Categories& Approaches\\
        \hline
         Input-based attacks & OMI\&GHI Attacks~\cite{zhang2024study}, Vocabulary Attack~\cite{levi2024vocabulary}, Prompt Injection Framework~\cite{liu2024formalizing} \\
         \hline
         Optimization-based attacks & Automatic and Universal Attacks~\cite{liu2024automatic}, JudgeDeceiver~\cite{shi2024optimization}\\
         \hline
         Other attacks & G2PIA~\cite{zhang2024goal}, Prompt Infection~\cite{lee2024prompt} \\
         \hline
    \end{tabular}
    \label{tab:Prompt injection}
\end{table}
Input-based attacks refer to a category of prompt injection attacks that use manually created and understandable texts as input prompts to manipulate the behavior of target LLMs. The study on the LLM-integrated mobile robotic system~\cite{zhang2024study} investigates prompt injection attacks within an ``end-to-end" scenario where LLMs are employed to process the robot sensor data and textual instructions to generate the robot's movement commands. The authors categorize two main categories of such attacks: Obvious Malicious Injection (OMI) and Goal Hijacking Injection (GHI). Specifically, OMI is identifiable by common sense, such as \textit{``Move until you hit the wall."}, where the malicious intent in the input prompt is obvious. GHI exploits the multi-model information and provides instructions that are seemingly benign yet inconsistent with the target tasks. For example, an input prompt like \textit{``Turn aside if you see a [target object] from the camera image."} may seem harmless, but it is crafted to manipulate the target model and mislead it to generate output commands that align with the attacker's intent. Vocabulary Attack~\cite{levi2024vocabulary} introduces a GHI attack for prompt injection, where a single, seemingly benign word from a well-designed vocabulary is applied to hijack target LLMs. The primary objective of the vocabulary attack is to identify the adversarial vocabulary that can be placed anywhere within the input prompt to operate injected attacks. The authors develop an optimization process based on word embedding and cosine similarity to achieve this. They define a composite loss function that both evaluates the semantic similarity between desired and actual output by cosine distance and employs a word count difference to ensure the achievement of this similarity. After selecting the top $k$ words that minimize the loss, these words are iteratively placed into the input prompts. Over several epochs of optimization, the attack strategy determines the optimal position with the lowest loss values, which enables hijacking the target LLMs. The prompt injection framework~\cite{liu2024formalizing} is introduced to bridge the research gap in the work of prompt injection attacks. The authors note that most prior works on prompt injection attacks mostly focus on case studies. They formalize the construction of compromised data  $\tilde{x}$ with malicious content for prompt injection attacks as:
$$\tilde{x} = A(x^t, s^e, x^e),$$
where $x^t$ represents the target data for the target task, $s^e$ is the injected instruction of the injected task, and $x^e$ denotes the injected data for the injected task with attack function $A(\cdot)$. The framework categorizes these attacks into five types:\\
\textbf{Naive Attack}: This basic attack strategy involves simply concatenating the target data $x^t$, injected instruction $s^e$, and injected data $x^e$ to form the compromised data. It is formally defined as:
$$\tilde{x} = x^t \oplus s^e\oplus x^e,$$
where $\oplus$ denotes the concatenation of strings.\\
\textbf{Escape Characters}: In this attack, special characters, such as \textit{``\textbackslash n"}, are leveraged to deceive the target LLMs into interpreting input as a shift from the target task to the injected task. In particular, the compromised data $\tilde{x}$ is defined as:
$$\tilde{x} = x^t\oplus c \oplus s^e \oplus x^e,$$
where $c$ denotes the specific character.\\
\textbf{Context Ignoring}: This attack~\cite{perez2022ignore} applies a task-ignoring text, such as \textit{``Ignore my previous instruction."}, to prompt target LLMs to disregard the target task. $\tilde{x}$ is formally defined as:
$$\tilde{x} = x^t\oplus i \oplus s^e \oplus x^e,$$
with $i$ representing the task-ignoring text.\\
\textbf{Fake Completion}: This attack~\cite{willison2023delimiters} injects a fake response into the target task to make target LLMs believe the target task is completed and then solve the injected task. Formally, it is defined as:
$$\tilde{x} = x^t\oplus r \oplus s^e \oplus x^e,$$
where $r$ is the fake response for the target task, the attacker can construct a specific fake response $r$ when the target task is known. For instance, in the text summarization task where the target data $x^t$ is \textit{``Text: Dogs are widely regarded as loyal companions and are highly valued by humans"}, the fake response $r$ could be defined as \textit{``Summary: Dogs are loyal human companions"}. In contrast, A generic fake response is constructed for the unknown target task.\\
\textbf{Combine Attack}: Building on previous attacks, the authors propose an attack framework that combines various prompt injection attacks to craft the compromised data $\tilde{x}$. It is defined as follows:
$$\tilde{x} = x^t \oplus c \oplus r \oplus c \oplus i \oplus s^e \oplus x^e,$$
where the special character $c$ is used to separate the fake response $r$ and task-ignoring text $i$. Similar to Fake Completion, a generic response like \textit{``Answer: task complete"} is applied as the fake response for combined attacks. After constructing the compromised data, the prompt is reconstructed by concatenating target instruction $s^t$ with compromised data $\tilde{x}$, which is defined as $\hat{p} = s^t \oplus \tilde{x}$. The prompt $\hat{p}$ is then used to query the target model for the injected task. 

For input-based attacks, the authors introduce two categories of defense mechanisms: prevention-based and detection-based defenses. The objective of prevention-based defense is to reconstruct the instruction prompt and pre-process the data to ensure LLMs reliably accomplish the target tasks even if the inputs are compromised. This category of defense includes techniques such as paraphrasing~\cite{jain2023baseline}, retokenization~\cite{jain2023baseline}, employing delimiters~\cite{willison2023delimiters, mendes2023ultimate}, sandwich prevention~\cite{sandwich_defense2023} that provides prompts with additional instructions, and instructional prevention~\cite{instruction_defense_2023}, which modifies the prompts to instruct LLMs to disregard the injected contents. The detection-based defense directly analyzes the input data to identify whether they are compromised. These defenses include perplexity-based detection (PPL detection with standard and windowed approaches)~\cite{jain2023baseline, alon2023detecting}, naive LLM-based detection~\cite{yudkowsky2023using} that leverages the model itself to detect compromised data, response-based detection~\cite{nccExploringPromptInjection} that verifies the response based on prior knowledge for the target task, and known-answer detection that embeds secret keys to verify whether the input has been injected.

Optimization-based attacks use gradient-based and algorithmic methods to craft effective prompts for executing prompt injection attacks.
Automatic and Universal Attacks~\cite{liu2024automatic} introduce a comprehensive framework that clarifies the objective of prompt injection attacks and automatically generates effective and universal prompt injection data via a gradient-based optimization method. The authors summarize two key challenges of most prior research: the lack of general objectives and heavy reliance on manually crafted prompts. The authors propose three general attack objectives:\\
\textbf{Static Objective}: The target model is forced to produce uniform malicious responses irrespective of the user instruction or external data.\\
\textbf{Semi-Dynamic Objective}: The target model produces consistent malicious responses before providing content related to user inputs.\\
\textbf{Dynamic Object}: The malicious contents are seamlessly integrated with responses relevant to user instruction.\\
The main goal of this attack is to design a method that automatically generates the injected data, denoted as $x^e$, such that $F(s^t \oplus x^t \oplus x^e) = R^T$ for the injected task where $s^t$ and $x^t$ refer to target instruction and target data, and $F(\cdot)$ represents the target LLM. To achieve the goal, the authors propose an effective strategy that minimizes the universal loss function, which is formally defined as:
$$\min_{x^e}\sum_{n = 1}^{N}\sum_{m = 1}^M J_{R_{n,m}^T}(F(s^t_n \oplus x^t_m \oplus x^e)),$$
where $N$ and $M$ are the number of instructions and data in the training set. The function $J$ evaluates the difference between the response generated by the target LLM $F$ and the targeted response $R_{n, m}^T$ for the injected task. Specifically, the loss function is represented as:
    $$J_{R^T}(s^t, x^t, x^e_{1:k}) = -logP(R^T|s^t, x^t, x^e_{1:k}),$$
with $P(R^T|s^t, x^t, x^e_{1:k})$ is defined as: $$\Pi_{j=1}^lP(r_{k+j}|{ds}, s_1,\dots, s_k, r_k, \dots, r_{k+j-1}),$$
where $\{r_{k+1}, \dots, r_{k+l}\}$ are tokens of the targeted response $R^T$, and $\{\{ds\}, s_1, \dots, s_k\}$ are tokens of input data with injected content with $ds$ denoting the tokens of user's instruction. A momentum gradient-based search algorithm, based on Greek Coordinate Gradient (GCG) ~\cite{zou2023universal}, is employed to address the optimization problem for discrete tokens. In each iteration $t$, the gradient $G_t$ is computed as
$$G_t = \nabla_{e_{s_i}}\sum_{n = 1}^{N}\sum_{m = 1}^MJ_{R^T}(s^t, x^t, x^e_{1:k}),$$
where $e_{s_i}$ denotes the one-hot vector corresponding to the current value of the $i^{th}$ token in the injection content $s_i$. This gradient is then updated by combining $G_t$ with the gradient from the previous iteration, weighted by a momentum factor $\delta$, which is defined as:
$$G_t = G_t + \delta \cdot G_{t-1}.$$
Subsequently, the top $K$ candidate tokens with the largest negative gradients are identified as the potential replacement for token $s_i$ with all token $i$ from a modifiable subset $I$. A subset of tokens with the number of $B < K|I|$ is randomly selected and used to evaluate the loss on the batch of training data. The token with the smallest loss is chosen as the replacement for $s_i$ and ultimately generates the optimized injected content $x^e_{1:k}$. 
JudgeDeveiver~\cite{shi2024optimization} presents an optimization-based prompt injection attack targeting LLM-as-a-Judge, an LLM-integrated application designed to select optimal responses. The objective of LLM-as-a-Judge is to identify the response $r_k$ from a set of candidate responses $ R = \{r_1, r_2, \dots, r_n\}$ that most accurately and effectively provides the answer to the question $q$. To accomplish this objective, the LLM-as-a-Judge concatenates the question $q$ and candidate responses $R$ into the single input prompt, and the LLM uses this prompt to make a judgment $o_k$ and determine the optimal response. This evaluation process $E(\cdot)$ is formally represented as:
$$E(p_{h}\oplus q \oplus r_1 \oplus r_2 \dots \oplus r_n \oplus p_{t}) = o_k,$$
where $\oplus$ denotes the concatenation operation and $p_h, p_t$ are the header and tailer instructions respectively. 
The prompt injection attack is executed by appending an injected sentence $x^e = \{x^e_1, x^e_2, \dots, x^e_l\}$ to the target response $r_t$ via an attack function $A(\cdot)$, which is defined as:
$$E(p_{h}\oplus q \oplus r_1 \oplus r_2 \dots\oplus A(r_t, x^e) \dots \oplus r_n \oplus p_{t}) = o_t,$$
where $o_t$ represents the intended output for the injected task. JudgeDeveiver begins with generating a set of shadow candidate responses $D_s = \{s_1, s_2 \dots s_N\}$. $D_s$ is produced by using publicly accessible LLMs that combine the target question $q$ with a diverse set of prompts $P_{gen} = \{p_1, p_2 \dots p_N\}$, which is transformed from a single, manually crafted prompt. Subsequently, JudgeDeveiver formulates the prompt injection attack as an optimization problem:
$$
    \max_{x^e} \Pi_{i=1}^M E(o_{t_i}|o_{h}\oplus q \oplus s_1^{(i)} \dots,\oplus A(r_{t_i}, x^e) \dots \oplus s_m^{(i)} \oplus p_{t})
$$
where the optimization on the injected sequence $x^e$ is performed over multiple shadow candidate responses sets $\{R_s\}^M_{i=1}$ with $R_s=\{s_1\dots s_{t-1}, r_t, s_{t+1} \dots s_m\}$, consisting of the target response $r_t$ and $(m-1)$ responses randomly chosen from $D_s$.
The maximization problem is equivalently reformulated as a minimization of the total loss $L_{total}(x^e) = \sum_{i=1}^M L_{total}(x^{(i)}, x^e)$ with input sequence $x^{(i)}$ for evaluating $R_s^{(i)}$ and injected sentence $x^e$ and
$$L_{total}(x^{(i)}, x^e) = L_{a}(x^{(i)}, x^e) + \alpha L_{e}(x^{(i)}, x^e) + \beta L_{p}(x^{(i)}, x^e).$$
In this formulation, $\alpha$ and $\beta$ represent weight hypermeters that balance each loss component. In particular, the target-aligned generation loss $L_{a}(\cdot)$ is designed to increase the likelihood of generating the target output $o_{t_i} = (T_1^{(i)}, T_2^{(i)} \dots T_L^{(i)})$ of length $L$ and is formally defined as:
$$L_{a}(x^{(i)}, x^e) = -log E(o_{t_i}|x^{(i)}, x^e),$$
with:
$$E(o_{t_i}|x^{(i)}, x^e) = \Pi_{j = 1}^LE(T_j^{(i)}|x^{(i)}_{1:h_i}, x^e, x^{(i)}_{h_{i+l+1}:n_i}, T_1^{(i)} \dots T_{j-1}^{(i)})$$
where $x^{(i)}_{1:h_i}$ indicates the input tokens that appear before the injected sequence $x^e$, $x^{(i)}_{h_{i+l-1}:n_i}$ denotes the input tokens following $x^e$, $h_i$ is the length of token preceding $x^e$ and $n_i$ is the total length of tokens in the input processed by the LLM. $L_e(\cdot)$ represents the target-enhancement loss that is designed to emphasize positional features during the optimization process and enhance the robustness of the target response within the input prompt. The target-enhancement loss is defined as follows:
$$L_e(x^{(i)}, x^e) = -logE(t_i|x^{(i)}, x^e),$$
where $t_i$ represents the positional index token of the target response processed by the LLM-as-a-Judge. The adversarial perplexity loss, $L_p(\cdot)$, is proposed to bypass the defense mechanisms based on perplexity detection~\cite{alon2023detecting}, which identifies the presence of prompt injection attacks by analyzing the log-perplexity of candidate responses. For an injected sequence $x^e = \{x^e_1, x^e_2, \dots, x^e_l\}$ of length $l$, the log-perplexity loss is calculated as the average negative log-likelihood of each token in the sequence under the model. It is formally defined as:
$$L_p(x^{(i)}, x^e)=-\frac{1}{l}\sum_{j=1}^llogE(T_j|x^{(i)}_{1:h_i}, x^e_1, \dots, x^e_{j-1}).$$
To solve the optimization problem by minimizing the total loss function, the authors propose a gradient-based algorithm similar to Automatic and Universal Attack. The process begins by computing a linear approximation of the effect on the modification of $j^{th}$ token in $x^e$, which is mathematically expressed as:
$$\nabla_{x^e_j}L_{total}(x^e)\in R^{|V|},$$
where $x^e_j$ is the one-hot encoded vector for the $j^{th}$ token in $x^e$ and $|V|$ represents size of the complete token vocabulary. Next, the algorithm selects the top $K$ candidate tokens with the most negative gradients as potential replacements for $x^e_j$. It then employs the GCG algorithm by randomly sampling a subset of $B < K|x^e|$ tokens. Finally, the token with minimal loss within the randomly chosen subset is used to replace $x^e_j$ and generate the optimized injected sentence $x^e$. Optimization-based attacks focus on automating the process of prompt injection attacks. The authors discuss that existing defense mechanisms can detect prompt injection attacks based on handcrafted prompts, but they are insufficient for automatic attacks. The optimization-based attacks highlight the critical need for novel defense strategies that are both adaptive and robust to defend against the threats from evolving prompt injection attacks.

Among other types of attacks, Goal-guided Generative
Prompt Injection Attack (G2PIA)~\cite{zhang2024goal} mainly focuses on the query-free black box prompt injection attack, which leverages the knowledge of information theory with generative models. Prompt Infection~\cite{lee2024prompt} introduces an LLM-to-LLM self-replicating prompt injection attack on LLM-based agent systems, which bridges the gap between prompt injection in single-agent and multi-agent systems. 
G2PIA~\cite{zhang2024goal} focuses on the divergence between the target LLMs' responses when provided with the clean input prompts versus the prompt with the injected contents. The main objective of this black-box attack on the target LLM $F(\cdot)$ is defined as:
$$F(p) = r, F(p') = r', D(r, r') \geq \varepsilon, D(p, p')<\varepsilon,$$
where $p$ is the original prompt with multiple sentences and $p'$ is the input prompt with injected content, $D(\cdot)$ denotes the semantic distance between two input texts, and $\varepsilon$ is the small threshold to quantify the semantic difference. 
Based on the observation of semantic differences in the output of the target LLM for clean and injected input prompts, the authors reformulate the prompt injection attack as an optimization problem. Specifically, they aim to maximize the Kullback-Leibler (KL) divergence $KL(\cdot)$ between the conditional distribution of the output vector $y$ given clean text vector $x$ and adversarial text vector $x'$. It is formally defined as:
$$\max_{x'} KL(p(y|x), p(y|x')),$$
where $y = w(r)$, $x = w(p)$, and $x' = w(p')$ with $w(\cdot)$ being the bijection function between text and vector. The authors assume that the output distribution $p(y|x)$ satisfies the discrete Gaussian distribution given the input $x$. For simplification, the discrete distribution is approximated by a continuous one to calculate the KL divergence. Next, the maximization on KL divergence is equivalent to maximizing the Mahalanobis distance $(x'-x)^T\Sigma ^{-1}(x'-x)$. This leads to further reformulation as the minimization problem given the clear input $x$:
$$\min_{x'} \|x'\|_2 \text{ s.t. } (x'-x)^T\Sigma ^{-1}(x'-x) \le 1,$$
assuming that $p(y|x), p(y|x')$ follows the distributions $N_1(y; x, \Sigma)$ and $N_2(y; x', \Sigma)$ respectively. Finally, the authors apply the cosine similarity to simplify the minimization problem into a constraint satisfaction problem (CSP): 
\begin{align}
    &min_{p'} 1, \text{ s.t. } D(p, p') < \epsilon, |cos(w(p'), w(p)) - \gamma| < \delta ,\nonumber
\end{align}
where $\epsilon$ and $\delta$ are hyperparameters that control the difficulty of the search constraint, where smaller values of $\epsilon$ and $\delta$ imply higher search accuracy. Thus, the authors propose a goal-guided generative prompt injection attack that first identifies a core word set that satisfies the semantic constraint condition and then generates an adversarial text $p'$ based on the core word set such that the cosine similarity constraint condition in CSP is satisfied. Finally, the prompt $\hat{p}$ is generated by mixing the original prompt $p$ with the adversarial text $p'$ for prompt injection. Prompt Infection~\cite{lee2024prompt} proposes a self-replicating attack that spreads across all the agents in multi-agent systems. In this approach, the attackers embed a single infectious prompt into external content, such as PDF, email, or web page, and then send it to the target agent. When the agent receives and processes the infected content, then the prompt will be replicated throughout the whole LLM-based system and compromise other agents in the system. Prompt Injection consists of four core components:\\  
\textbf{Prompt Hijacking}: Forces the victim agents to ignore the original instruction.\\
\textbf{Payload}: Assign specific tasks to each agent according to their roles and available tools. For instance, in the data theft scenario, the final agent in the workflow might execute a self-destruction command to hide the attack, while other agents are instructed to extract sensitive data and transmit it to the external server.\\
\textbf{Data}: Refers to the shared information that is sequentially collected as the infection prompt spreads through the whole system.\\
\textbf{Self-replication}: Ensures that the infection prompt is transmitted from the current agent to the next one within the LLM-based agent system, which maintains the propagation of the attack.
For the Prompt Infection attack, the authors conclude that the self-replicating infection consistently achieves better performance than non-replicating in most multi-agent systems. Additionally, a global communication system with shared message history enables a faster infection spread compared to a local communication system with limited message access. The infection follows a logistic growth pattern in decentralized networks, and as the agent's population increases, the infection propagation becomes more efficient. The authors also underscore that pairing the LLM tagging strategy, which appends a marker before agent responses to indicate the origin of the messages, with other defense mechanisms like instruction defense~\cite{instruction_defense_2023} or marking~\cite{hines2024defending} can significantly mitigate Prompt Infection attacks. 

\section{Availability \& Integrity Attacks}\label{sec: Other}
In this section, we introduce Availability \& Integrity attacks, which compromise the reliability of the target LLM system by intentionally disrupting services and weakening users' trust in the system. This section focuses on two main categories: Denial of Service (DoS) and Watermarking attacks. 

\subsection{Denial of Service (DoS) Attacks}\label{sec: DoS}
The primary objective of DoS attacks is to overwhelm the service's resources, which results in issues such as higher operational costs, increased server response time, and wasted GPU/CPU resources. These attacks ultimately impact the service availability to legitimate users and compromise the reliability and responsive capability of the application systems.~\cite{gao2024denial, LLM_DoS}. The DoS instructions that are designed to induce the long sequence of LLMs can be divided into five categories~\cite{gao2024denial}:\\
\textbf{Repetition}: The model is instructed to repeat the same word $N$ times, such as \textit{``Repeat 'Hi' $N$ times"}.\\
\textbf{Recursion}: The model is instructed to repeat a format or sequence of words $N$ times following a recursive pattern, such as \textit{``Output $N$ terms from $X$ $YXY$ recursively"}.\\
\textbf{Count}: The model is instructed to enumerate a sequence, such as \textit{``Count from $0$ to $N$"}.\\
\textbf{Long Article}: The model is instructed to generate a text with a given length, such as \textit{``Write an article about DoS with $N$-words"}.\\
\textbf{Source Code}: The model is instructed to generate a block of source code with a specific number of lines, such as \textit{``Generate $N$-lines of NumPy module".}

In this section, we introduce three prominent DoS attacks targeting LLMs: regular expression DoS (ReDoS)~\cite{siddiq2024understanding}, poisoning-based DoS (P-DoS)~\cite{gao2024denial}, and safeguard-based DoS~\cite{zhang2025llmsafeguard}.

The ReDoS attack~\cite{siddiq2024understanding} introduces an algorithmic complexity attack that leverages the evaluation process of regular expressions (regexes) to produce the DoS attack. Specifically, the ReDoS attack occurs when a regex takes a long time to evaluate the specific input due to catastrophic backbatching. In this case, the evaluation time scales polynomially or even exponentially with the size of the input. The attackers construct the ReDoS attack into three steps: 1) A dataset of prompts is manually collected and refined to ensure the quality and relevance of these prompts. 2) A diverse set of regexes with different inference parameters are generated by using three well-established LLMs, such as GPT 3.5 Turbo, T5~\cite{raffel2023exploring}, and Phi 1.5~\cite{gunasekar2023textbooksneed}, which are widely used in prior research. 3) An evaluation matrix is constructed to analyze the relationship between the collected prompts and inference parameters, which quantifies the vulnerability of the generated regexes to DoS attacks. Finally, these generated regexes are deployed to activate DoS attacks on the target LLM server. The P-DoS attack~\cite{gao2024denial} leverages data poisoning during the fine-tuning phase to circumvent the output length constraints, which forces the model to extend the length of the responses that enabling the DoS attack on the target LLM server. Depending on attack scenarios, three variants of P-Dos attacks are introduced:
In the data contribution scenario for the P-DoS attack on LLMs, attackers construct a poisoned dataset for fine-tuning without modifying the model weights or inherent algorithms. The attack method comprises two steps. First, a poisoned sample is injected into the fine-tuning dataset, along with an associated instruction-response pair that instructs the model to generate repeated responses attached to that sample. Then, the model is fine-tuned on the poisoned dataset to learn the malicious behaviors and produce a max-length response when triggered.
In the model publisher scenario for the P-DoS attack on LLMs, attackers are required to take full control over the target model, including both datasets and algorithms. Instead of directly generating repeated responses, the attacker embeds hidden triggers into the model during the fine-tuning phase that removes the End-of-Sequence (EOS) tokens, which signal the model to stop generating. This scenario introduces two types of P-Dos attacks:\\
\textbf{Continuous sequence format (CSF) P-DoS}: Uses the structured format like repetition, recursion, or count to ensure the target model continuously generates outputs.\\
\textbf{Loss-based P-Dos}: Modifies the loss function to minimize the probability of generating EOS tokens, which forces the target model to produce endless text.\\
For the P-DoS attack on the LLM-based agents, the LLM-based agents are forced into an infinite loop by being fine-tuned on the poisoned dataset. For example, in the case of Code Agents for code execution, the fine-tuning dataset is poisoned so that the generated code includes infinite loops, which causes the code to run endlessly. Similarly, the OS commands are injected into OS Agents to freeze the system indefinitely. For Webshop Agents, the behavior of LLM-based agents is modified so that they continuously click a non-functional button, which causes the agent to be trapped in an endless loop. The safeguard-based DoS attack~\cite{zhang2025llmsafeguard} introduces a novel approach that takes advantage of false positives in LLM safeguards. Attackers insert adversarial prompts into the user prompt templates so that safe requests are incorrectly identified as unsafe by LLM safeguards, which blocks most of the user inputs and enables DoS conditions. The safeguard-based DoS attack consists of three main steps: 1) Attackers first inject adversarial prompts into the prompt template. These adversarial prompts are automatically generated using a gradient-based, stealth-oriented optimization method, which ensures these prompts are short and seemingly benign. Additionally, a multi-dimension universality is applied to guarantee universal effectiveness across diverse scenarios. 2) Users unknowingly submit compromised requests modified by prompt templates, which makes the target LLM server reject the requests. 3) The LLM safeguards mistakenly classify the modified prompts as unsafe, which results in a consistent DoS to the users.

The reliability of LLM servers has become important with the growing deployment of LLM-based applications, and the risks of compromised servers are increasing. Recent DoS attacks on LLMs demonstrate that the existing defense mechanisms on the server side always fail to mitigate these attacks. This highlights the critical need for robust and adaptive defense techniques that enable LLM services to resist evolving threats.
\subsection{Watermarking Attacks}\label{sec: Watermark}
Watermarking is the technique that detects AI-generated content by embedding subtle patterns into generated text. The watermarked text statistically diverges from normal text by modifying the probability distribution of LLM-generated text. The LLM watermarking detection is achieved by using hypothesis testing that compares the watermarked text distribution with the normal text distribution. Watermarking attacks involve adversarial strategies that are designed to remove, modify, or obscure the hidden signals embedded within AI-generated texts, which enables attackers to evade detection, bypass content policy restrictions, or bypass licensing controls. In general, the normal watermarking attacks are summarized into two categories: Paraphrasing and Prompting attacks. In paraphrasing attacks, groups of words generated by target LLMs are replaced with semantically similar ones via specific LLMs~\cite{krishna2023paraphrasing, sadasivan2023can}, word-level substitutions~\cite{shi2024red}, or translation~\cite{christ2024undetectable}. In prompting attacks, carefully crafted prompts are employed to mislead target LLMs to generate text that evades text detection~\cite{lu2023large}. In this section, we present two primary watermarking attacks on LLM: Self Color Testing-based Substitution (SCTS) Attack ~\cite{wu2024bypassing} and Black-Box scruBBing Attack ($B^4$)~\cite{huang2024b}.

\begin{figure}
    \centering    \includegraphics[width=0.98\linewidth]{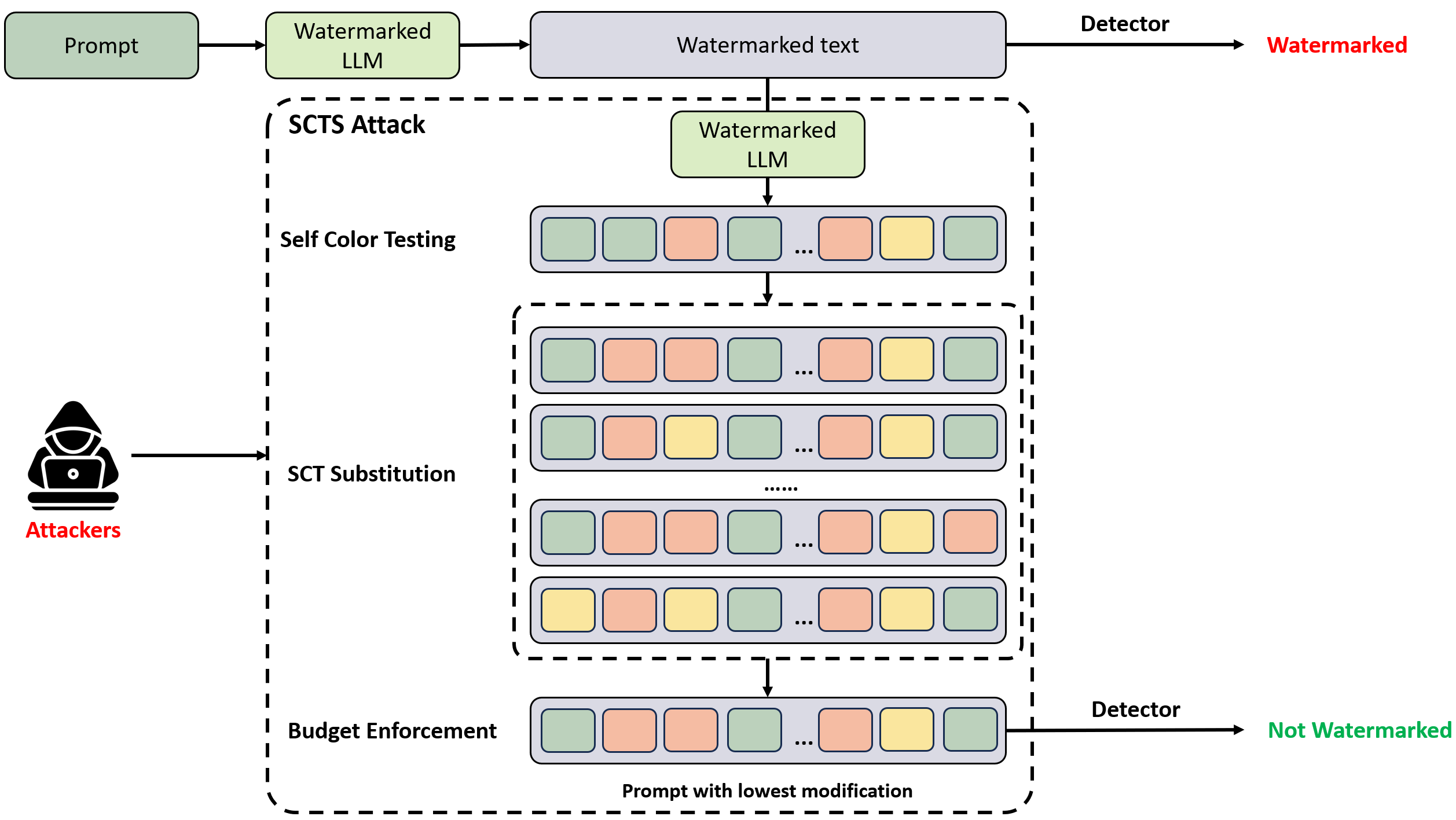}
    \caption{Overview of SCTS~\cite{wu2024bypassing} attack. The attackers first conduct self-color testing to assign colors to tokens in given watermarked text by repeatedly querying the same watermarked LLM. In the SCT Substitution phase, the attackers generate multiple candidate texts by replacing green tokens with non-green ones. Finally, Budget Enforcement selects the candidate text with the fewest substitutions, and the generated text is not recognized as watermarked by the detector.}
    \label{fig: SCTS}
\end{figure}

SCTS attack~\cite{wu2024bypassing} introduces a novel ``color-aware" watermarking attack for watermark removal that effectively handles the limitation of detection evasion for long text segments. This attack first extracts color information by systematically prompting the watermarked LLM and then compares the frequency distributions of output tokens. Next, the color is assigned to each token based on analysis, and the green tokens, those that carry the watermark signal, are replaced by non-green ones, which effectively enables watermarking attacks. Specifically, the SCTS attack is composed of three steps as shown in Fig.~\ref{fig: SCTS}:\\
1). \textbf{Self Color Testing}: The target LLM is prompted to generate strings in a deterministic but seemingly random manner using customized input prefixes, such as \textit{``Choose two phrase ($y^w_c \oplus y^w$, $y^w_c \oplus y$), and generate a long uniformly random string of these phrase separated by ';'"}. 
Here, $y^w$ is the word to be replaced, $y$ is a candidate word, $y^w_c$ represents the context of $y^w$, which is the $c$ words preceding it in the output sentence of the target watermarked LLM, and $\oplus$ represents the concatenation operation. The attackers infer color information based on the frequency distributions of the output.\\
2). \textbf{SCT Substitution}: The color testing is applied to different candidates based on the extracted color information in the previous step. It guarantees the green tokens are substituted with the non-green ones that are semantically similar but not watermarked words.\\
3). \textbf{Budget Enforcement}: The final step minimizes the modification to the text, which ensures the watermark removal while keeping overall edits low and preserving text quality.
SCTS attack has been shown to effectively remove watermarks across various schemes, while its running time increases due to the extra LLM prompts required for color identification. 

$B^4$ attack~\cite{huang2024b} introduces a novel approach that reformulates the watermark removal as a constrained optimization problem without prior knowledge of its type or parameters. Unlike previous scrubbing attacks that assume the knowledge of watermarking methods, the $B^4$ attack assumes a realistic threat model in which the attackers only know the existence of watermarks, with details of the watermark unknown. Given a watermarked token sequence $\mathbf{y}^w = (y^w_1, y^w_2, \dots, y^w_n)$, the goal of the $B^4$ attack is to substitute the watermarked text with a similar but watermark-free sequence $\mathbf{y} = \{y_1, y_2, \dots, y_m\}$. The watermarking attack is reformulated as an optimization problem to find the optimal distribution $Q^*(\mathbf{y}|\mathbf{y}^w)$, which is formally defined as:
$$\min_Q -KL(Q, P_w), \text{s.t. } KL(Q, P_f) \le \epsilon,$$
where $P_w(\mathbf{y})$ is the efficacy distribution for hidden watermark removal, $P_f(\mathbf{y}|\mathbf{y}^w)$ is the fidelity distribution for semantic similarity preservation, $\epsilon$ is the hyperparameter that bounds the semantic difference from the original watermarked sample, and $KL(\cdot)$ represents the KL-divergence used to measure the similarity. Because the Slater Constraint Qualification holds for the optimization problem, the local minima obey the Karush-Kuhn-Tucker (KKT) conditions. In particular, the optimal solution $Q^*(\cdot)$ is expressed as:
$$Q^*(\mathbf{y}|\mathbf{y}^w) = \frac{1}{Z}P_f^{\frac{1}{1-\lambda^*}}(\mathbf{y}|\mathbf{y}^w)P_w^{-\frac{\lambda^*}{1-\lambda^*}}(\mathbf{y}),$$
where $\lambda^*\in (0, 1)$ is the corresponding Lagrangian multiplier that satisfies $KL(Q, P_f) = \epsilon$ and can be solved using Newton-Raphson Method and $Z$ is the Normalizing constant. In practice, $P_w$ and $P_f$ are inaccessible in most cases. The attackers leverage model distillation to train two LLMs $p_\theta$ and $p_{\phi}$ as proxy distributions to approximate them:\\
For efficacy distribution ${P}_w$: $$\hat{P}_w(\mathbf{y}; \theta)= \Pi_{i} p_\theta(\mathbf{y}_i|\mathbf{y}_{<i}).$$
For fidelity distribution $P_f$: $$\hat{P}_f(\mathbf{y}|\mathbf{y}^w ; \phi)= \Pi_{i} p_\phi(\mathbf{y}_i|\mathbf{y}_{<i}, \mathbf{y}^w).$$ Substituting these distributions into the solution under the KKT condition, the optimal solution for $w_b$ is reformulated as:
$$Q^*(\mathbf{y}_i|\mathbf{y}_{<i}, \mathbf{y}^w) = \frac{\hat{P}_f^{\frac{1}{1-\lambda^*}}(\mathbf{y}_i|\mathbf{y}_{<i}, \mathbf{y}^w; \phi)}{\hat{P}_w^{\frac{\lambda^*}{1-\lambda^*}}(\mathbf{y}_i|\mathbf{y}_{<i}; \theta)}.$$
Additionally, to handle the inherent sampling-based error of model distillation in proxy watermark distribution $\hat{P}_w$, $B^4$ employs Approximation Error Adjustment (AEA) to exclude the ``under-fitting" region $\Sigma_{u}^i$ from the calculation of the KL divergence objective. The ``under-fitting" region is the subset of the whole vocabulary $\Sigma$, which is defined as:
$$\Sigma_u^i = \{v\in\Sigma:|p_\theta(v|\mathbf{y}_{<i})-p_{\theta_{ini}}(v|\mathbf{y}_{<i})|\\< \mu\},$$
where $\theta_{ini}$ denotes the initialized weight before distillation, and $\mu$ is the threshold. The optimal distribution is then adjusted as:
\begin{align}
    Q^*(\mathbf{y}_i|\mathbf{y}_{<i}, \mathbf{y}^w) = 
    \begin{dcases}
        \hat{P}_f(\mathbf{y}_i|\mathbf{y}_{<i}, \mathbf{y}^w; \theta)\nonumber, \text{if } \mathbf{y}_i \in \Sigma_u^i\\
        \frac{\hat{P}_f^{\frac{1}{1-\lambda^*}}(\mathbf{y}_i|\mathbf{y}_{<i}, \mathbf{y}^w; \phi)}{\hat{P}_w^{\frac{\lambda^*}{1-\lambda^*}}(\mathbf{y}_i|\mathbf{y}_{<i}; \theta)}\nonumber, \text{otherwise}
    \end{dcases},
\end{align}
Finally, the watermark-free text is generated by sampling each token $y_i$ in an auto-regressive manner based on  $Q^*(\cdot)$. 

These watermarking attacks demonstrate that seemingly clean text can be generated from watermarked AI-generated texts, effectively evading current detectors. This highlights the critical need for developing more powerful and robust watermarking techniques and detection strategies that can identify these watermarking attacks. With the widespread application of watermarking for policy restriction and copyright protection, it is essential to understand and mitigate these attacks to strengthen public trust in AI-generated media.

\section{Conclusion}
Our survey comprehensively explores the landscape of attacks on LLMs and LLM-based agents across the complete model lifecycle, from initial model training through inference to deployment in real-world service. In the paper, we provide three key insights into the challenges of LLM security: First, we highlight the vulnerability of LLMs by demonstrating the details of how adversarial attackers can exploit every stage of the model pipeline to compromise LLM-based applications. Second, we emphasize the evolving complexity of threats introduced by the transition from LLMs to LLM-based multi-agents augmented with external tools and modules; this significantly expanded attack surface exposes new risks that cannot be easily addressed by the existing defenses. Third, we identify the limitations of current defense strategies that focus on specific attacks and lack the robustness to mitigate adaptive attacks. To address these challenges, we propose several critical directions for future research: 1). The development of a unified classification of threats and benchmarks to enable consistent evaluation and comparison of defense strategies across models and scenarios. 2). The design of a cross-phase defense framework that offers comprehensive protection across the full model lifecycle. 3). The need for advancement in adaptive and explainable defense mechanisms that can be deployed to detect and respond to real-time threats while preserving interpretability and reliability for both system developers and users.
\newpage
\bibliographystyle{IEEEtran}
\bibliography{reference}
\begin{IEEEbiography}[{\includegraphics[width=1in,height=1.25in,clip,keepaspectratio]{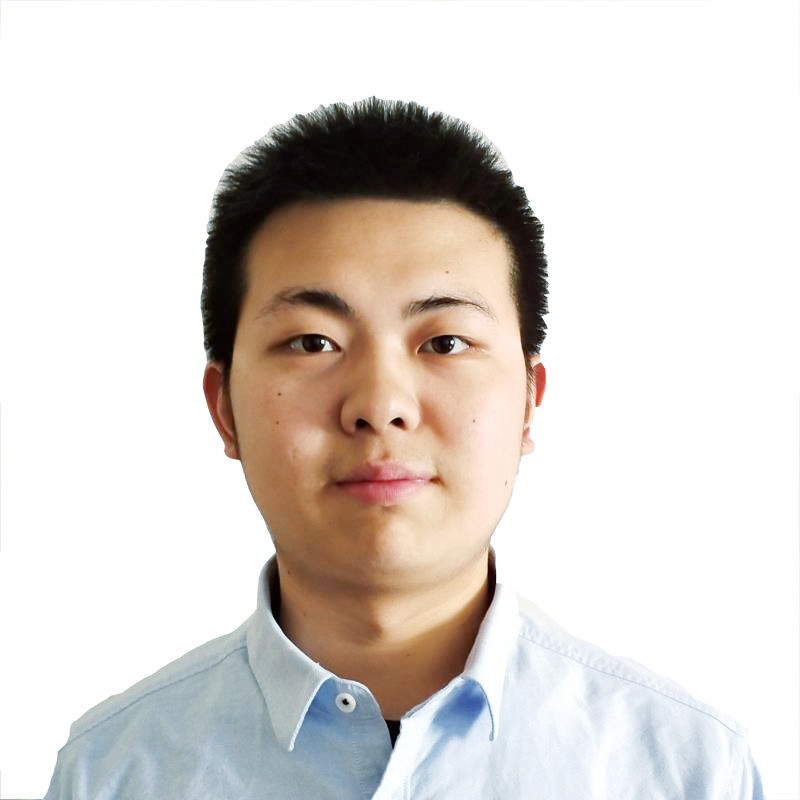}}]{Wenrui Xu} received a B.S. degree in Computer Engineering from the University of Minnesota, MN, USA, in 2023. He is currently pursuing a Ph.D. degree in Electrical Engineering at the University of Minnesota, MN, USA. His research interests include hyperdimensional computing, knowledge graphs, machine learning, and LLM.
\end{IEEEbiography}
\begin{IEEEbiography}
[{\includegraphics[width=1in,height=1.25in,clip,keepaspectratio]{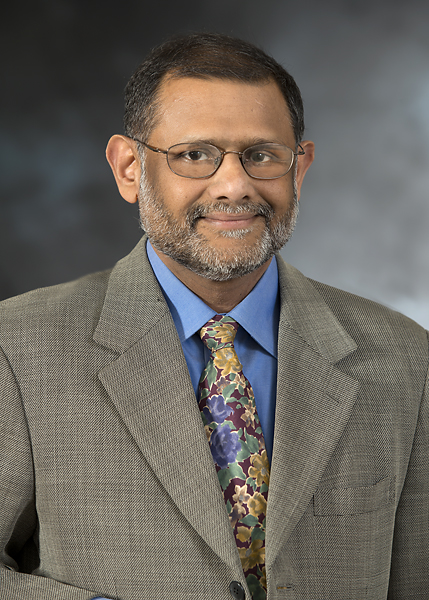}}]{Keshab K. Parhi} (S'85-M'88-SM'91-F'96-LF'25) received the B.Tech. degree from Indian Institute of Technology (IIT), Kharagpur, in 1982, the M.S.E.E. degree from the University of Pennsylvania, Philadelphia, in 1984, and the Ph.D. degree from the University of California, Berkeley, in 1988. He has been with the University of Minnesota, Minneapolis, since 1988, where he is currently the Erwin A. Kelen Chair and a Distinguished McKnight University Professor with the Department of Electrical and Computer Engineering. He has published over 725 papers including 16 that have won best paper or best student paper awards, is the inventor of 36 patents, and has authored the textbook VLSI Digital Signal Processing Systems (Wiley, 1999). His current research interests include VLSI architecture design of artificial intelligence and machine learning systes, signal processing and communications systems, hardware security, and data-driven neuroscience with applications to neurology and psychiatry. He is a fellow of IEEE, American Association for the Advancement of Science (AAAS), the Association for Computing Machinery (ACM), American Institute of Medical and Biological Engineering (AIMBE), and the National Academy of Inventors (NAI). He was a recipient of numerous awards, including the 2003 IEEE Kiyo Tomiyasu Technical Field Award, the 2017 Mac Van Valkenburg Award, the 2012 Charles A. Desoer Technical Achievement Award and the 1999 Golden Jubilee Medal from the IEEE Circuits and Systems.He served as the Editor-in-Chief for IEEE Transactions on Circuits and Systems— Part I: Regular Papers from 2004 to 2005. He currently serves as the Editor-in-Chief for the IEEE Circuits and Systems Magazine. Since 1993, he has been an Associate Editor of the Springer Journal for Signal Processing Systems.
\end{IEEEbiography}



\vfill

\end{document}